\documentclass[11pt]{article}

\usepackage{mathptmx,graphicx}
\usepackage[it]{caption2}

\captionwidthfalse
\renewenvironment{abstract}
  {\begin{list}{}{\addtolength{\itemsep}{0.2cm} \leftmargin 0.5in \rightmargin 0.5in
           \itemindent \listparindent
           \rightmargin \leftmargin
           \item\relax}\small}{\vspace{0.0cm}\normalsize\end{list}}
\renewcommand{\title}[1]{
  \begin{center}
  \Large\textsf{\textbf{#1}}
  \end{center}\vspace{0.0cm}
  \normalfont\normalsize
}
\renewcommand{\author}[1]{
  \begin{list}{}{\leftmargin 0.5in \rightmargin 0in
           \item\relax}\raggedright #1\end{list}
}
\newcommand{\address}[1]{
  \begin{list}{}{\leftmargin 0.5in \rightmargin 0in
           \item\relax}\raggedright\textit{#1}\end{list}
}
\newcommand{\sect}[1]{\vspace{0.9cm} \textsf{\textbf{#1}}\newline\par}
\renewcommand{\section}[1]{}
\renewcommand{\subsection}[1]{\vspace{0.9cm} \textit{\textbf{#1}}\newline\par}

\newcommand{\fm}{F_{\rm Maxw}}
\newcommand{\ra}{\langle r \rangle}

\begin{document}

\title{Characterizing the intermediate phases through topological analysis}

\author{Mykyta V. Chubynsky}
\address{Department of Physics, University of Ottawa, 150 Louis Pasteur,
Ottawa (Ontario) K1N 6N5, Canada}

\begin{abstract}
I review computational studies of different models of elastic network
self-organization leading to the existence of a globally isostatic (rigid but unstressed) or nearly isostatic
intermediate phase. A common feature of all models considered here is that
only the topology of the elastic network is taken into account; this allows
the use of an extremely efficient constraint counting algorithm, the pebble
game. In models with bond insertion without equilibration, the intermediate
phase is rigid with probability one but stress-free; in models with
equilibration, the network in the intermediate phase is maintained in a
self-organized critical state on
the verge of rigidity, fluctuating between percolating and nonpercolating but
remaining nearly isostatic. I also consider the connectivity analogs of these
models, some of which correspond to well-studied cases of loopless percolation
and where another kind of intermediate phase, with existing but nonpercolating
stress, was studied.
\end{abstract}

\sect{Introduction}
The use of rigidity considerations and in particular, constraint counting,
in describing the composition dependence of properties of chalcogenide glasses
goes back to the 1979 paper of J.C.~Phillips~\cite{phillips79}. It had been realized
at least since the time of Kirkwood~\cite{kirkwood} that the main interactions in covalent
materials are those between first neighbors [central-force (CF) or bond-stretching
constraints] and those between second neighbors [angular or bond-bending (BB)
constraints] in the covalent network; all other forces are much weaker and can
be neglected in the first approximation. If one counts these strong covalent
constraints and subtracts their number from the number of degrees of freedom (d.o.f.) in
the system ($3N$ in a system of $N$ atoms), then one gets an approximation to
the number of zero-frequency motions in the system, as first proposed by
J.C.~Maxwell~\cite{maxwell} who considered stability of engineering constructions. These
zero-frequency modes acquire a small nonzero frequency once the neglected
weaker forces are added and are referred to as {\it floppy modes}. I will
denote their number in the system $F$ and the Maxwell approximation to this
number obtained by counting constraints as described above (the procedure called
{\it Maxwell counting}), $\fm$. Given the network, $\fm$ is, of course, very
easy to calculate and Maxwell suggested that it can serve as an estimate of the
overall rigidity of the network. A completely rigid network in three
dimensions has $F=6$ zero-frequency motions (3 translations and 3 rotations).
If $\fm>6$, then, assuming that $\fm$ is a good approximation to $F$ in this
case, the network will have some additional, internal floppy motions besides
the rigid-body motions, and will be {\it flexible} --- the higher $\fm$, the more
flexible the network. If $\fm<6$, then, since rigid-body motions are always
present and thus $F<6$ is impossible, $\fm$ is clearly incorrect; this,
however, is still useful information in that it indicates that there are more 
constraints than needed to make the network rigid and so some fraction of
the constraints can be deemed {\it redundant}. If, in a thought experiment,
constraints are added one by one, some of them will be inserted into an
already rigid region of the network where all the distances are already fixed, and
{\it generically}, that is, unless these constraints happen to have very
special lengths, they will become strained and will introduce {\it stress}
into the network; such a network will be {\it rigid and stressed}. Finally, at
$\fm=6$ the network has just enough constraints to be rigid, but not
overconstrained. For a large network, such as a bulk glass, the number of rigid
body motions (6) can be neglected compared to the total number of d.o.f.
($3N$) (especially given that these considerations are approximate
anyway) and the boundary between the rigid and flexible networks is assumed to
correspond to $\fm=0$. Note that at this point the number of constraints
balances the number of d.o.f. For a glass network, $\fm$ will
depend on its composition (namely, on fractions of atoms with different
valences), and Phillips~\cite{phillips79} realized that glasses with compositions
corresponding to $\fm$ are the best glassformers. Indeed, they are rigid and
so cannot explore their configurational space and find the global, crystalline
energy minimum as easily as flexible glasses can; yet, they are also
stress-free and so being disordered is not too energetically costly. Since
Phillips' work, extrema of several other physical quantities were associated
with the same point $\fm\approx 0$ (for a list, see, e.g., Ref.~\cite{travbool}).

Four years after Phillips' work, Thorpe~\cite{thorpe83} realized that the transition
between flexible and rigid networks can be viewed as the
{\it rigidity percolation} transition. In ordinary {\it connectivity}
percolation~\cite{stauffer}, one likewise deals with a disordered network of sites connected by
links and asks whether there exists a connection between the
opposite sides of the network. This can be equivalent to asking if there
exists a connected fragment of the network (a {\it cluster}) that spans the
network; in the limit of an infinite network ({\it thermodynamic limit}), such a
spanning (or {\it percolating}) cluster will be likewise infinite. If
links are conducting and a voltage is applied between the opposite sides of the
network, then there will be a current whenever a percolating cluster exists
(a percolating network), otherwise, there will be no current. Rigidity
percolation is very similar. One now considers networks of elastic springs
(elastic networks) and defines {\it rigid clusters} as parts of the network that
behave as rigid bodies (i.e., distances between all sites remain fixed) in all
possible motions that do not deform the springs. One then asks whether there is
a percolating rigid cluster spanning the network. The analogs of the voltage and
current are strain and stress: if the network is strained, this will introduce
additional stress into the network and will increase its elastic energy if and
only if rigidity percolation occurs. (There are some subtleties that will
become apparent later in this review.) Just as connectivity percolation, rigidity
percolation is expected to be a {\it phase transition}, with an infinitely
sharp threshold between nonpercolating and percolating networks, so that when
the bonds are randomly removed, in the thermodynamic limit all networks with
the bond concentration (or the mean coordination number) below a certain value are
nonpercolating and all networks with the bond concentration above this value are
percolating. Besides rigidity percolation, one can also consider
{\it stress percolation}, by defining {\it stressed regions} as contiguous
regions of the network where all bonds are stressed and asking if
any such region percolates. Based on Maxwell counting, one would expect the
rigidity and stress percolation thresholds to coincide, since it is only
at a single point, corresponding to $\fm=0$, that the network should be
rigid without being stressed. As we will see, this may or may not be the
case in reality.

The view of the rigidity transition as a percolation phase transition was almost
immediately confirmed in numerical simulations~\cite{fengsen}. Detailed studies of this
transition were, hovewer, very difficult. In connectivity percolation,
the effective conductance of a network of conductors may depend on the physical
details, such as the resistance of each link; {\it but} the configuration of clusters and,
in particular, the existence of a percolating cluster only depend on the
{\it topology} of the network (i.e., what sites are connected to what sites),
and so finding clusters for a particular network
and determining whether it percolates is very easy computationally. It is not
immediately obvious if the same is true for rigidity percolation, i.e., if
the rigidity of a particular piece of a network is determined entirely by its
topology or also depends on the details such as the spring constants and the
equilibrium lengths of the springs. In any case, it was not known how to test
for rigidity using just the network topology. Maxwell counting would be exactly
such a test, but it is known to be approximate, because the presence of
redundant constraints can never be ruled out; besides, in its original form it
only determines the rigidity of the network as a whole and cannot, e.g.,
decompose the network into rigid clusters. For this reason, early studies of
rigidity percolation had to resort to the computationally costly procedure of
choosing a physical realization of the network, with all spring lengths and
force constants, and either relaxing it under strain or diagonalizing its
dynamical matrix. Only small networks, up to a few hundred sites, could be
studied.

The breakthrough came with the involvement of mathematical rigidity
theory~\cite{servatius,graver,whiteleyphysbio}.
It turned out that some rigidity properties of the network, such as the
number of floppy modes and the configuration of rigid clusters and stressed
regions, are indeed determined entirely by the network topology, {\it but}
only for {\it generic} networks. That is, among all possible realizations of
a given topology with different spring lengths, etc., nearly all have the same
rigidity properties, {\it except} for an infinitesimally small fraction of
{\it nongeneric} networks. Such nongeneric networks have something special
about them, such as the presence of parallel bonds or more than two bonds whose continuations
intersect at the same point. Approaches to studying rigidity that were
subsequently developed are only valid for generic networks and only such
networks are considered in what follows in this review. This means that the
results would not necessarily apply to periodic networks, even randomly diluted.
If a network topologically equivalent to a periodic network (such as a crystal
lattice) is considered, one has to make an additional assumption that the
network is ``distorted'' by introducing a disorder in spring lengths. But for
fully disordered models of amorphous solids we should be safe, even when some
of the bond lengths are the same (as is always the case in real systems).

A theorem by Laman~\cite{laman} enabled the construction of a very fast algorithm, the
{\it pebble game}~\cite{jacobs95,jacobs97}, for analyzing the rigidity of a network based on its
topology. The essence of the theorem is that by applying Maxwell counting to all
{\it subnetworks} of the network, along with the network as a whole, one can
find all redundant constraints and thus obtain the correction to the Maxwell
count of floppy modes. I
describe the pebble game algorithm in more detail in the next section.

The pebble game enabled much more detailed studies of the rigidity transition
using much larger networks. Randomly diluted networks were studied first, with
either bonds (bond dilution) or sites with all associated bonds (site dilution)
removed at random (with some subtleties, as explained below). In this case,
both the rigidity and stress transitions were found to be continuous (second
order) transitions and the rigidity and stress thresholds were found to be the
same. Especially careful studies were carried out for CF networks
in 2D, where critical exponents and thresholds (for both bond and site dilution
on the triangular lattice) were found~\cite{jacobs95,duxbury95,jacobs96}. While it is not immediately
obvious that 2D networks are good models of 3D glass networks, in most cases it
is found that the results are qualitatively the same, and, of course, the
advantage of 2D models is that networks of much larger linear sizes can be
considered computationally, reducing the finite-size effects. For this reason,
many of the models described in this review were considered in 2D, and we will
be switching between 2D and 3D throughout.

Note that while the original idea by Phillips justified the existence of
optima of various quantities at the point where constraints and degrees of
freedom balance, the rigidity percolation approach predicts thresholds and
singularities, such as those associated with other kinds of phase transitions.
The optima are expected to be rather robust against the introduction of weaker
interactions neglected in the elastic network model of a glass. However,
singularities can be blurred by such interactions (as well as entropic effects
at a finite temperature~\cite{joos}) and for this reason are much
harder to observe in real systems. Nevertheless, both extrema and thresholds in
various quantities have indeed been observed (for a review, see, e.g.,
Ref.~\cite{travbool}). To much surprise, however, Boolchand {\it et al.} found that
in many cases, {\it two} distinct transitions are observed. Since the first
hints of this in Ge-Se~\cite{travbool} and a more definite observation in Si-Se glasses~\cite{bool99sise}, this has been found in
many other cases, as reviewed by P.~Boolchand in this volume. Perhaps the most
striking are the results for the non-reversing heat flow
integrated across the glass transition, as measured by modulated differential
scanning calorimetry (MDSC). It was observed that in many cases, there exists a
very broad region in the composition dependence of this quantity where
it is very low and almost constant, with a very sharp rise outside the
region on both sides. Other anomalies, e.g., in vibrational frequencies, are
observed at the boundaries of this region. These observations suggested that
instead of a single optimal point in the composition phase diagram (Phillips'
view) or a single threshold (Thorpe's view), a whole region of finite width is
optimal and thresholds exist on both sides of this region, which could not
be explained by theory.

Two ideas were perhaps key in further theoretical developments. First, it was
immediately suggested by Boolchand {\it et al.}~\cite{travbool} that some features of
medium-range structure were perhaps responsible for the double transition.
Medium-range order was, of course, ignored in studies of rigidity percolation,
where randomly diluted networks were considered, but always exists in reality,
since some local structures are less energetically favorable and thus less
likely; this had to be incorporated into theory in some way. Second, if
``optimal'' glasses are indeed those that are rigid but unstressed, then a
broad minimum perhaps means that such rigid-unstressed glasses exist in a
region of finite width, rather than at a single point. This is indeed plausible,
since a glass network would try to {\it self-organize} to avoid unnecessary
stress while remaining rigid. The question was
how to incorporate these features in a model without making it too complicated.
Ideally, one would like to reproduce the medium-range structure of the glass as
faithfully as possible. This is very hard to do from first principles, although
some work in this direction has been done (see the paper by Inam {\it et al.} in
this volume). Another possibility is using atomistic modeling with empirical
potentials, as described in the contribution by Mauro and Vashishta. Such
simulations are still too time-consuming and the potentials are not always
reliable. Instead, Thorpe {\it et al.}~\cite{thorpe00} asked if it is possible to stay
entirely within the purely topological pebble game approach. As a reminder, from
the network topology, using the pebble game, one can find whether stress is
present in the network, but not the stress energy (if stress does exist). In
the original model of network self-organization by Thorpe {\it et al.}, the
network is constructed one bond at a time. Candidate bonds for insertion are
selected at random (as in the case of random bond dilution), but for each
candidate the pebble game is used to find whether stress is created, and all
bonds producing stress are rejected. At some point, finding a place where
insertion would not create stress becomes impossible, and from that point on,
insertion is continued at random. Since the creation of stress is delayed,
the stress transition occurs at a higher bond concentration than the rigidity
transition, and the rigid-unstressed {\it intermediate phase} forms in between,
in qualitative agreement with experiments. In the third section of this review,
we describe the studies that have been performed using variants of
this model in two and three dimensions, as well as a model with additional
medium-range order and a connectivity analog, {\it loopless percolation}, where
certain aspects of self-organization and the
intermediate phase are easier to understand. Also, I describe a variant of the
model leading to another kind of intermediate phase where stress occurs but
does not percolate. This was studied for connectivity, but should also exist in
the rigidity case.

Further work on topological models of self-organization was motivated by the
following observation. At least
within the purely topological approach, there is no reason to give preference
to any stress-free network structure over any other stress-free structure, and
it is reasonable to assume that they are all equiprobable, i.e., what is known
as the {\it uniform
ensemble} of networks should be generated. This actually
corresponds to thermodynamic equilibrium, i.e., the microcanonical ensemble (or
the canonical ensemble at zero temperature) of networks in the model where all
stress-free networks have the same energy, but all stressed networks have a
higher energy. (Of course, this reasoning neglects weak forces and the
possibility that different networks have different entropies, but there is no
way to take this into account within the topological approach.) The problem is
that in the model in which the bonds are only added and never removed, different
networks are not equiprobable. This is similar to how an aggregation process
cannot lead to the equilibrium structure. This reasoning led to the introduction
of the process of network {\it equilibration}. The resulting model of
{\it self-organization with equilibration} is considered in the fourth section
of this review. This leads to a variant of the intermediate phase somewhat different
from the original one: there is still no stress, but rigidity only percolates with a
finite probability. However, the network always stays very close to being
{\it isostatic} (rigid but unstressed) as a whole and is in a state resembling
{\it self-organized criticality} (SOC)~\cite{bak}, although, unlike common SOC, this
is observed in an equilibrium system.

Finally, in the fifth section I consider some open questions concerning the
existence of the equilibrated intermediate phase at nonzero temperatures and
whether the variant of the intermediate phase with stress but no stress
percolation survives equilibration.

\pagebreak

\sect{Maxwell counting and the pebble game}

As mentioned, the Maxwell counting approximation $\fm$ for the number of floppy modes
is the difference between the number of d.o.f. (which is $dN$ for a network of
$N$ sites in $d$ dimensions) and the number of constraints $N_c$:
\begin{equation}
\fm=dN-N_c.
\end{equation}
This neglects the presence of redundant constraints. Since adding a redundant
constraint to the network does not change the number of floppy modes $F$, the
{\it exact} result for $F$ is
\begin{equation}
F=dN-N_c+N_R,\label{Fexact}
\end{equation}
where $N_R$ is the number of redundant constraints. The problem is how to find
$N_R$.

I first give the results of Maxwell counting in
some cases of interest and then describe its exact extension, the pebble game
algorithm, that allows the evaluation of $N_R$ and thus obtaining the exact
value of $F$.

\subsection{Maxwell counting results}
As mentioned, self-organization models were
considered both for 2D CF networks and 3D BB glass
networks. Let us consider the simpler 2D case first. In this case, only
CF constraints are present, so only first neighbors are connected by
constraints and the number of constraints is the same as the number of bonds.
We introduce the mean coordination number $\ra$ as the number of bonds
connecting a site averaged over all sites. If there are $N$ sites in the
network, then, since each bond is shared between two sites, the total number of
bonds (and thus constraints) is $N\ra /2$. The number of d.o.f. is
$2N$ and so
\begin{equation}
\fm =2N-N\ra /2.\label{max2D}
\end{equation}
This becomes zero at $\ra=4$, so we expect the rigidity transition to be
at $\ra_c^{\rm 2D}\approx 4$. Given this result, in order to study rigidity
percolation in 2D by bond dilution, we need a lattice with the coordination
number higher than 4, and the triangular lattice is a natural choice.

The case of 3D glass networks is somewhat more subtle. We are interested in
modeling chalcogenide glasses containing atoms of valence 2 (chalcogens S, Se,
Te) and possibly 3 (As, P) and/or 4 (Ge, Si). The possible inclusion of halogens
(valence 1) has to be treated separately, for reasons that will be clear in a
moment. It is assumed that the coordination number (the number of neighbors) of
an atom always coincides with its valence. Each atom of valence $r$ has
$r$ associated CF constraints, each of them is shared with another atom and so
will enter with the factor of 1/2 in the total constraint count. The total
number of angular constraints associated with the atom is $r(r-1)/2$, which
gives 1, 3, 6 constraints for $r=2$, 3, 4, respectively. However, for an atom
of valence 4 only 5 out of 6 angular constraints are independent, even for an
isolated atom with its neighbors. This means that generically, the
sixth constraint creates stress, however, this stress is unavoidable, since no
changes in the structure can get rid of it, and we ignore it in what follows.
Obviously, counting only 5 constraints out of 6 improves the result for the
number of floppy modes, and this is commonly done since Thorpe's work~\cite{thorpe83},
even though technically this means going beyond Maxwell counting. The number of
(independent) angular constraints can then be written as $2r-3$. This indeed
produces 1, 3, 5 for $r=2$, 3, 4 (it turns out it is even valid for $r>4$,
although we do not need this). For $r=1$, however, this produces $-1$, a
meaningless result (it should be zero, as there are no angular constraints
associated with a 1-fold coordinated site). This is why presence of halogens
requires a separate consideration. If the number of atoms is $N$ and the average
coordination number is $\ra$, the total number of constraints is
\begin{equation}
N_c=N[\ra/2+(2\ra-3)],
\end{equation}
and since the number of d.o.f. is $3N$,
\begin{equation}
\fm = 3N-N_c=N(6-5\ra/2).\label{fm3D}
\end{equation}
This becomes zero at $\ra=2.4$, so the conclusion is that the rigidity
transition should occur at $\ra_c^{3D}\approx 2.4$. Note that the Maxwell
result for the number of floppy modes, Eq.~(\ref{fm3D}), depends on the
mean coordination $\ra$ only and not on other details of the composition. If
atoms of valence 1 are also present, this is no longer the case. The threshold
in this case also depends on the fraction of 1-coordinated atoms $n_1$:
Maxwell counting predicts $\ra_c\approx 2.4-0.4n_1$~\cite{boolthorpe}, although this is not a good approximation when there are many bonds between atoms of 
valence 1 and 2~\cite{boolpriv,mythesis}.

Finally, it is interesting to note that connectivity problems can be
viewed as rigidity problems with one d.o.f. per network site,
regardless of the dimensionality of the network (with the resulting number of
floppy modes $F$ equal to the number of connected clusters). This means, in particular,
that Maxwell counting can also be applied to connectivity percolation. With
just first-neighbor constraints (links), the number of constraints is $N\ra/2$;
the number of d.o.f. is $N$; so
\begin{equation}
\fm = N-N\ra/2.
\end{equation}
This is zero at $\ra=2$, so $\ra_c^{\rm con}\approx 2$. For the bond-diluted
square lattice with coordination 4, this corresponds to the bond fraction
$p_c=\ra_c^{\rm con}/4\approx 1/2$, which happens to be the exact result in
this case~\cite{stauffer}. For the triangular lattice (coordination 6),
$p_c=\ra_c^{\rm con}/6$, and Maxwell counting gives $p_c\approx 1/3$, which is
very close to the exact result
$p_c=2\sin(\pi/18)\approx 0.347$~\cite{stauffer}. For the honeycomb lattice
(coordination 3), the Maxwell counting result is $p_c\approx 2/3$, likewise very
close to the exact result $p_c=1-2\sin(\pi/18)\approx 0.653$~\cite{stauffer}.

\subsection{The pebble game}
As mentioned, Maxwell counting is approximate, because it does not take into
account redundant constraints. Besides, even in cases when the floppy mode
count is correct ($F=\fm$), another potential source of problems is that even
an overall rigid (i.e., percolating) network can have floppy inclusions and thus
the number of floppy modes can still be larger than zero (or the rigid body
count). Also, since Maxwell counting is an overall (mean-field) count, studying,
e.g., the details of the structure of rigid clusters is impossible. The pebble
game algorithm fixes all these problems.

The first question to ask is: since the errors in the floppy mode count are due
to redundant constraints, how can we detect them? Obviously, if Maxwell counting
gives the number of floppy modes $\fm$ that is smaller than the number of rigid
body motions $N_{RB}$ [3 in 2D, 6 in 3D, 1 in 1D, i.e., for connectivity, or in general,
$d(d+1)/2$ in $d$ dimensions], then redundant constraints must be present. This, however, does not
detect all instances, since it is possible that redundant constraints coexist
with floppy modes and if there are more of the latter than of the former,
then the Maxwell count will be more than $N_{RB}$. However, it may still be
possible to detect redundancy, if the counting is done not for the network as
a whole, but for a {\it subnetwork} that includes the stressed region containing
redundant constraints, but not the floppy regions. The question is if this is
always possible, i.e., if it is true that for every network containing
redundant constraints, there exists a subnetwork for which Maxwell count is
less than $N_{RB}$. In 2D, the answer is yes, and this statement is known as
the Laman theorem~\cite{laman}. In 3D, unfortunately, there are exceptions, since it
is possible to have networks with redundant constraints and floppy modes
residing inseparably in the same region, e.g., the infamous
{\it double-banana graph}~\cite{servatius}.
However, it is {\it conjectured} that such situations do not happen in
{\it bond-bending networks}, i.e., networks that always have a second-neighbor
BB constraint between any two adjacent first-neighbor constraints, and glass
networks are exactly of this kind. This is known as the {\it molecular
framework conjecture}~\cite{tay,whiteley,whiteleyphysbio}. While it has not been proved, it has not been
disproved either in 20 years of extensive studies, and we feel safe to rely on
it when studying glass networks. As an aside, it was shown
recently~\cite{3DCF} that
it is also safe to use the same approach for 3D {\it central-force} bond- and site-diluted networks, as errors, although possible, are extremely rare.

Note that the Laman theorem and its 3D analog, the molecular framework
conjecture, only give a way to find out whether redundant constraints are
present in a given network, but not their number in case they are present. The
solution is to insert network constraints one by one, starting from the ``empty''
network (all sites, no bonds), and test each newly inserted constraint for
redundancy, by doing constraint counting for all subnetworks including the new
constraint. When all constraints are inserted, the count of redundant
constraints is obtained.

\begin{figure}
\begin{center}
\includegraphics[width=10cm]{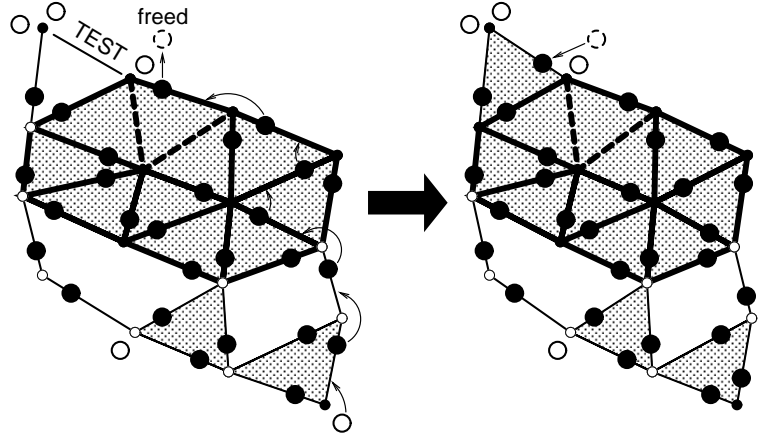}
\caption{An illustration of the 2D pebble game. Small circles (open or filled)
are sites, lines are bonds connecting them. There are two pebbles per site,
some free (large open circles) and some covering bonds (large filled circles).
The goal is to free 4 pebbles at the ends of the test bond. Three pebbles are
free initially, and the fourth one can be freed through a sequence of pebble
movements, as shown by arrows in the left panel. During these movements, all
covered bonds remain covered, but can be covered from either side. Since the
freeing process succeeds, the test bond is independent and is covered by one
of the four free pebbles at its ends. After this, 4 free pebbles remain, thus
the network has 4 floppy modes, including 3 rigid body motions. Two bonds
(thick dashed lines) were found redundant previously, as the corresponding pebble searches
failed, and are not covered. The associated failed search region is the region
of stress (thick bonds). Rigid clusters larger than a single bond are shaded.
Sites shared between several rigid clusters (pivots) are small open circles,
all other sites are small filled circles. Adapted from Refs.~\cite{jacobs95,czech01}.\label{pebble}}
\end{center}
\end{figure}

Even given the above approach to counting redundant constraints, it is still
not immediately clear how to implement it efficiently, since the number of
subnetworks in a large network is huge (exponential in the network size). In
order for the approach to be useful, one needs a method to test all
subnetworks simultaneously. A way to do this was suggested by
Hendrickson~\cite{hendrickson92}
and the resulting algorithm is known as the {\it pebble game}~\cite{jacobs95,jacobs97,jacobs98,travbio,3DCF}. The idea is to
match constraints to d.o.f. by associating d.o.f. with
{\it pebbles} that can cover constraints. In 2D, in particular, two pebbles are
assigned to each site (Fig.~\ref{pebble}), so the total number of pebbles is $2N$ and is equal to
the number of d.o.f. During the pebble game, a pebble can be free
or cover a constraint. Initially, there are no constraints, so all pebbles are
free. Constraints are inserted one by one and each constraint is tested for
redundancy by attempting to collect four free pebbles at its ends while keeping
all covered constraints covered. If this is possible, this guarantees that for any
subnetwork including the new constraint, the difference between the number of
d.o.f. and the number of constraints is at least four without the
new constraint, thus at least three when it is added, which, according to the
Laman theorem, means that the new constraint is independent. Constraints deemed
independent are covered with one of the four free pebbles, those deemed
redundant are not covered and are ignored in subsequent considerations as far
as the redundant constraint count is concerned. If the four pebbles at the end
of a constraint are not already free, freeing of a pebble is possible, if there
is a free pebble at the other end of one of the constraints connecting the
site where freeing is attempted, in which case the free pebble covers the
constraint and the covering pebble is freed. Sometimes this swapping procedure
may have to be repeated, if the free pebble is found not at a neighbor of the
site where a free pebble is desired, but at a neighbor of a neighbor, etc. A
pebble retrieval in such a case is illustrated in Fig.~\ref{pebble}. If freeing
the fourth pebble fails, the region over which the failed search proceeded is
the region of stress induced by the new constraint. When all constraints are
added, the exact count of redundant constraints, and thus, according to
Eq.~(\ref{Fexact}), the exact number of
floppy modes $F$, is obtained. In fact, it is easy to see that $F$ is equal to
the number of free pebbles. Likewise, stressed regions are known. Decomposition
into rigid clusters requires a separate procedure, where a constraint is
selected, three pebbles are freed at its ends, and then the surrounding region
where it is impossible to free a pebble is a rigid cluster; this continues until
all rigid clusters are mapped. For more details of the pebble game procedure and
its justification, see Ref.~\cite{jacobs97}.

In 3D, the pebble game procedure is similar in spirit, but the straightforward
implementation is somewhat more complicated. There are, of course, three pebbles
per atom. One major difference compared to 2D is that after the maximum number
of pebbles (six) are freed at the ends of a newly inserted constraint, one
still needs to check whether the seventh pebble can be freed at all of the
neighbors of at least one of the ends. This procedure can in principle be used
for both BB and general (non-BB) networks~\cite{jacobs98,3DCF}, although in the latter case it is
not guaranteed to be correct in all cases. Even in the BB case, since
constraints are inserted one by one, the network cannot remain bond-bending at
all times, and in order for the pebble game to be correct, one needs to make
sure that after every insertion of a CF constraint, all associated BB constraints
are added immediately. This keeps the network as close to bond-bending as
possible, and, as part of the molecular framework conjecture, it is assumed that
this is sufficient for the pebble game to remain correct. Alternatively, one
can insert constraints in arbitrary order, but make sure that for any
4-coordinated site, only 5 out of 6 associated angular constraints are inserted.
A technically simpler variant of the 3D pebble game (but intrinsically only
applicable to BB networks) is based on equivalence between bond-bending networks
and {\it body-bar networks}~\cite{jacobs98,travthorpe}: if in a 3D bond-bending network, all sites
are replaced by bodies having (like any rigid bodies in 3D) 6 d.o.f.
and all {\it bonds} (i.e., CF constraints) are replaced by 5 bars
connecting the bodies replacing the sites the bond connects, with all angular
constraints eliminated, the resulting network has the same number of d.o.f.
and the same
rigid clusters and stressed regions as the original network (there
are some subtleties for 1-coordinated and disconnected sites that need to
be considered separately). Therefore, one can run a ``6-dimensional'' pebble
game assigning 6 pebbles per site, trying to free 11 pebbles at the ends of the
test bond, and if successful, covering the bond with 5 of the freed pebbles.
This is the version of the pebble game that is commonly used for 3D BB networks,
in particular, in the FIRST software for the rigidity analysis of
proteins~\cite{proteins01,FIRST}.

Finally, the pebble game can also be used for connectivity problems. Since
connectivity is equivalent to rigidity with one d.o.f. per site,
there should be one pebble per site; testing for redundancy is done by
attempting to free two pebbles at the ends of the link being tested. The
pebble game is comparatively less useful in connectivity than in rigidity,
since finding connected clusters is rather straightforward; still, in some
problems using the pebble game can give an advantage.

\sect{Self-organization without equilibration}
In this section, I describe the applications of the original approach to
self-organization developed by Thorpe {\it et al.}~\cite{thorpe00,czech01}. I treat the simpler
2D case first and then the more relevant case of 3D glass networks, which is
mostly similar to the 2D case, but with some technical subtleties. In both
cases, I start with a brief review of rigidity percolation studies with
random insertion, without self-organization, and then compare to the
self-organized case. I then discuss a newer work by
Sartbaeva {\it et al.}~\cite{asel07},
where the width of the intermediate phase is related to
the distribution of chain lengths in the network. I then describe the
connectivity analog of the model, which turns out to be a previously studied
variant of loopless percolation, although a new feature is the extension into
the stressed (``loopy'') phase where an interesting mean-field-like exactly
linear dependence of conductivity on link concentration was observed. We
finish the section by describing a different variant of the intermediate
phase, stressed but non-stress-percolating, that was studied for connectivity,
but should exist in the rigidity case, too.

\subsection{2D central-force networks}
\begin{figure}
\begin{center}
\includegraphics[width=8.5cm]{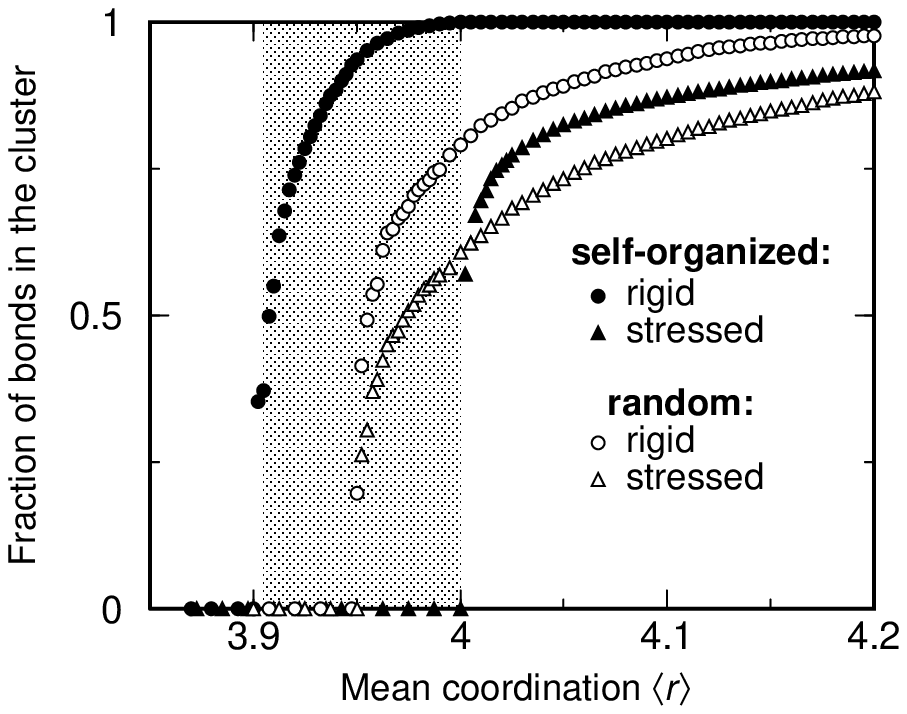}\vspace{0.2cm}
\includegraphics[width=8.5cm]{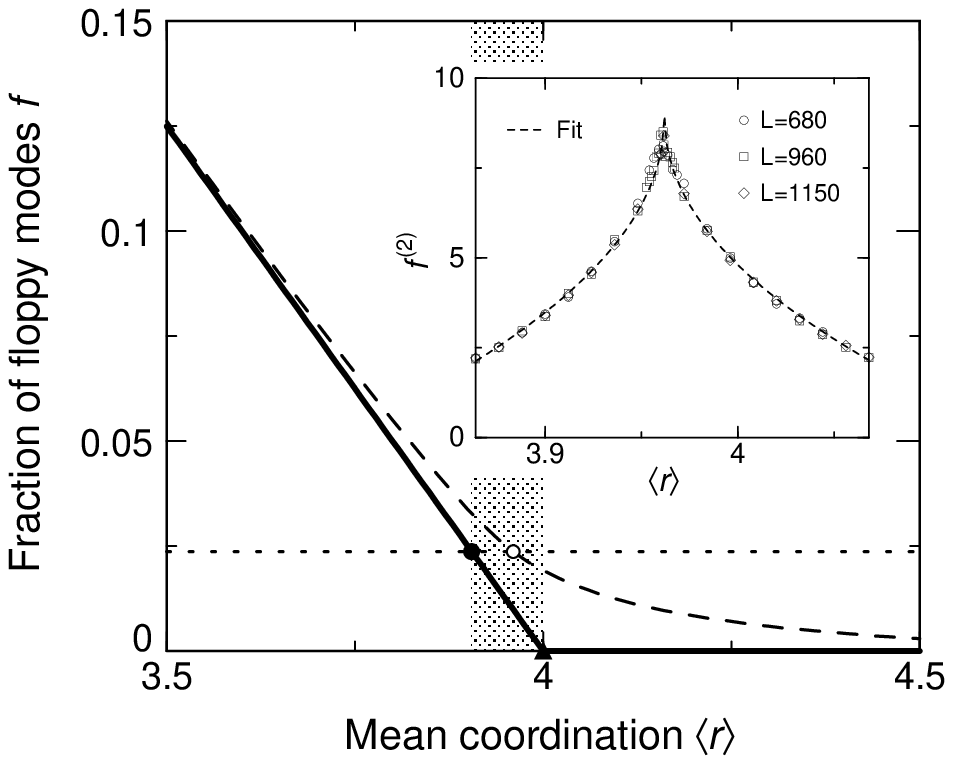}
\caption{The results of rigidity analysis for 2D central-force bond-diluted
networks built on the triangular lattice, both random and self-organized.
The figure is adapted from Refs.~\cite{jacobs96,czech01}.
(Top) The fractions of bonds in the percolating rigid cluster and the
percolating stressed region. In the random case, the rigidity and stress phase
transitions coincide; in the self-organized case, they do not, and the
intermediate phase forms in between (shaded). All results are averages over
two networks of $400\times 400$ sites.(Bottom) The number of floppy
modes per d.o.f., $f=F/2N$, in the random (dashed line) and self-organized
(solid line) cases. In the random case, the combined rigidity and stress
transition occurs at the point marked by the open circle; there is a cusp
in the second derivative at this point, as shown in the inset. In the
self-organized case, the rigidity and stress transitions are marked by filled
symbols (a circle and a triangle, respectively). In this case, the number of
floppy modes coincides with the Maxwell counting result up to the stress
transition, and there are no singularities, cusps, etc. at the rigidity
transition, but there is a break in the slope at the stress transition. The
intermediate phase in the self-organized case is shaded. Note that the rigidity
transitions occur at the same $f$ in the random and self-organized cases, as
explained in the text.
\label{2D}}
\end{center}
\end{figure}

Rigidity percolation on 2D networks obtained by {\it random} bond dilution was
first studied by Feng and Sen~\cite{fengsen} and then, once the pebble game became
available, in much more detail by Jacobs and Thorpe~\cite{jacobs95,jacobs96}. The latter
authors found that the rigidity transition on the randomly bond diluted
triangular lattice occurs at $\ra_c=3.961\pm 0.002$, very close to the Maxwell
counting prediction $\ra=4$. In percolation transitions, the fraction of
the network in the percolating cluster is commonly used as the order parameter:
indeed, it is zero in the nonpercolating phase, as only finite clusters exist
there, but becomes nonzero when the infinite cluster appears in the percolating
phase. In the top panel of Fig.~\ref{2D}, the fractions of bonds in the percolating rigid cluster and
in the percolating stressed region are plotted (open symbols). It is seen, first of all, that
the stress transition occurs {\it at the same point} as the rigidity transition;
thus there is only one, combined transition. Second, both quantities change
continuously, growing from zero starting at the threshold. Thus the transition
in this case is {\it continuous}, or {\it second order}. Another interesting
quantity is the number of floppy modes $F$. In connectivity percolation, when
it is viewed as rigidity percolation with 1 d.o.f. per site, the number of
floppy modes is simply the number of clusters; on the other hand, Fortuin and
Kasteleyn~\cite{fortkast72} showed that connectivity percolation is equivalent to a limit of
the Potts model~\cite{wu} when the number of states formally approaches 1; one
can therefore define the free energy and it turns out to be the negative number
of clusters. We can extend this result to rigidity percolation, assuming that
in this case as well, $-F$ is the free energy~\cite{dux99}. While there is no proof
similar to that for connectivity percolation, it was shown~\cite{dux99} that this
quantity has the correct convexity (the second derivative of $-F$ with
respect to the mean coordination is always negative, similar to how the
second temperature derivative of the free energy, which is the specific heat
with the minus sign, should always be negative).
The number of floppy modes $F$ is plotted in the bottom panel of Fig.~\ref{2D}
(dashed line). The
line looks smooth; however, if the second derivative is calculated (shown in the
inset of Fig.~\ref{2D}), there is a
cusp at the transition, very similar to the behavior of the specific heat at thermal
phase transitions, thus giving another confirmation of the role of $-F$ as the
free energy.

As already mentioned in the introduction, in the approach by Thorpe {\it et al.}
to network self-organization one starts with an empty lattice and inserts bonds
one by one rejecting those that would create stress. Testing of bonds for
whether they create stress is done by the pebble game; thus the pebble game
serves a dual purpose being used both for network construction and its analysis.
Once inserting bonds without creating stress becomes impossible, bond insertion
continues at random (it is obvious that if for a given network stressless
insertion is impossible, then it will be all the more impossible once some
extra bonds are added, so there is no need to check whether stressless insertion
becomes possible again at a later time). The results for the percolating
rigid cluster and the percolating stressed region obtained using this
approach are shown in the top panel of Fig.~\ref{2D} (filled symbols). The most striking difference
compared to the random (non-self-organized) case is that the rigidity and
stress transitions no longer coincide: a percolating rigid cluster appears below
the point where stress becomes inevitable, so there is an {\it intermediate
phase} which is rigid but unstressed. The lower boundary of the intermediate
phase (the rigidity transition) lies at $\ra\approx 3.905$. The upper boundary (the stress transition) coincides with the point where avoiding stress is no longer possible; at
that point, stress appears and immediately percolates. This upper boundary lies
at $\ra=4$, which is the Maxwell counting result for the rigidity transition.
This is not coincidental and, in fact, it can easily be shown that $\ra=4$
is the exact value for the point at which stress becomes inevitable. Indeed,
both in the floppy phase (below the rigidity transition) and in the intermediate
phase there is no stress, thus no redundant constraints, so the Maxwell
counting result for the number of floppy modes is exact ($F=\fm$). This is seen
in the bottom panel of Fig.~\ref{2D}, where the number of floppy modes in the self-organized case is
plotted as the solid line, and it is seen that below $\ra=4$ this is a
straight line that coincides with the Maxwell counting result
[Eq.~(\ref{max2D})]. At the point at which inserting a nonredundant constraint
is no longer possible, a constraint inserted in any place in the network will
be redundant and thus will not change the number of floppy modes. But, as
already mentioned, if a constraint is redundant and its insertion creates
stress, then it will still be redundant if it is inserted at any later time in
the bond insertion process. This means that once the point at which any
insertions would cause stress is reached, reducing the number of floppy modes
further is impossible, and thus at this point $F$ should equal 3 (the number
of rigid body motions), negligible in the thermodynamic limit. But since up to
this point Maxwell counting is correct, then this is the point at which
$\fm=0$, which is the Maxwell counting appoximation to the rigidity transition
point, or $\ra=4$ in our case. This consideration establishes the point at which
the switch from stress avoidance to random insertion occurs and stress appears;
it is still not obvious that (as happens to be the case here) stress immediately
percolates once it arises, and, in fact, we will see that this is not so in a
different model.

Note that the number of floppy modes does not exhibit any singularities at the
rigidity transition; on the other hand, at the stress transition, there is a
break in the slope. If $-F$ is interpreted as the free energy, a break in
the slope corresponds to a first order transition, however, there is no evidence
of a jump in the size of the stressed region at the stress transition. This may
mean that the interpretation of $-F$ as the free energy is no longer correct
when the network is not random.

Two other details about the results for self-organized networks are worth
mentioning. First, the fraction of the network in the percolating rigid cluster
is exactly 1 in the stressed phase (top panel of Fig.~\ref{2D}, filled circles), i.e., the whole network is in the
percolating cluster. This is obvious from the above consideration, since at
the point where stress appears no internal floppy modes are left, so the whole
network should be rigid. This is in contrast to the random case, where the
fraction of the network in the percolating cluster does not reach 1 until
full coordination, $\ra=6$. Second, at the rigidity percolation transition
the number of floppy modes is the same in the random and self-organized
cases (bottom panel of Fig.~\ref{2D}). This is not coincidental, and, in fact, is a corollary of a
more general correspondence between random and self-organized networks. Namely,
self-organized networks are constructed by trying bonds at random, like in the
random case, but rejecting redundant bonds. But redundant bonds do not
influence either the configuration of rigid clusters or the number of floppy
modes, so whether they are rejected or not should not influence these
properties. As a consequence, random and self-organized networks having the
same number of floppy modes should also have the same statistics of rigid
clusters and, in particular, the same fraction of the network in the percolating
cluster.

Elastic properties of the network in the intermediate phase are rather
interesting. Of course, to calculate the values of the effective elastic moduli,
the topology alone is not enough --- one needs to assign
spring lengths and force constants --- but for qualitative conclusions, these
details should not matter. A natural expectation is that below the rigidity
percolation transition, the elastic moduli (both the bulk modulus and the
shear modulus) should be zero, as the network can respond to external strain
by moving finite rigid clusters with respect to each other without any energy
cost, but above the transition, the moduli should be nonzero, as the percolating
cluster has to be deformed, which should cost some energy. For randomly diluted
networks, this is indeed the case~\cite{garboczi}. Surprisingly, in self-organized
networks, while the elastic moduli {\it may} be finite for finite networks in
the intermediate phase, they tend to zero in the thermodynamic limit and become
nonzero only above the stress transition. Moreover, in the particular case of
networks with periodic boundary conditions (PBC), the elastic moduli are exactly
zero even for finite networks! This seems outright impossible at first glance:
as stated above, the percolating cluster has to deform, after all, and
shouldn't there be an energy cost? In fact, there is no contradiction: by
definition, PBC imply that only motions such that all ``copies'' of the atom
in all supercells move in the same way are considered, and so when saying that
a particular region of the network is rigid, we consider only such motions.
However, when an external strain is applied, this is done by changing the size
and/or shape of the supercell, which moves the ``copies'' with respect to
each other, and so this motion is not from the ``allowed'' subset and what we
call the percolating rigid cluster should not be (and, in the intermediate phase,
is not) rigid with respect to such motions. That applying external strain to
a network with PBC in
the intermediate phase should not cost any energy is easy to see: since the
presence or absence of stress depends only on the network topology, changing
the size or shape of the supercell cannot introduce stress to a network that
was originally unstressed. Since in the thermodynamic limit the elastic moduli
should not depend on the boundary conditions, in this limit the moduli should
be zero for any boundary conditions. Of course, once the neglected weak
interactions and entropic effects are taken into account, the elastic moduli in the intermediate phase (as
well as in the floppy phase, for that matter) will no longer be zero. Still,
these results perhaps indicate that some sort of a threshold (albeit blurred)
or a crossover in the elastic properties should be apparent at the upper
boundary of the intermediate phase, but less so at the lower boundary.

\subsection{Glass networks}
The results of the self-organization model for glass networks are overall similar to
those for 2D CF networks, but there are some qualitative differences. These
differences arise for two reasons: first, the details of how the networks are
constructed, and second, the fact that there are several constraints per bond
(as insertion of a bond also adds the associated angular constraints to the
network).

As we have seen, Maxwell counting predicts that the rigidity properties of
3D BB networks should not depend much on the details of the composition, other
than the mean coordination, {\it but} only if there are no atoms of coordination
1 or 0, the presence of which shifts the transition downward very significantly.
For this reason, the case when such atoms are present should be modeled
separately and normally, even when talking about random bond dilution, one
implies {\it restricted} dilution where atoms of coordination 2, 3 and 4 only are
allowed. The studies of rigidity percolation on such networks were done by
starting with a network with all atoms 4-coordinated (one can use amorphous
Si models constructed by the WWW method~\cite{WWW} or its
improvements~\cite{WWWnew},
although even the topologically ordered diamond lattice turns out to be
adequate as well) and then diluting it at random, but with the restriction that
no atoms of coordination 1 appear. This is not a perfect way of constructing
random networks with the only restriction of no 0- and 1-coordinated atoms, as,
for example, the fractions of bonds between atoms of particular coordinations
differ from what they should be if the connections are fully
random~\cite{travthorpe,mythesis}.
However, these details should not matter much. Once a point well below
$\ra=2.4$ is reached, the resulting network is considered as the starting
point, the pebble game analysis is run for it and then bonds are put back one
by one while running the pebble game. This procedure leads to the rigidity
transition located at $\ra=2.375$ for the diamond lattice and $\ra=2.385$ for
the a-Si network obtained by the WWW procedure~\cite{travthorpe}, very close to the
Maxwell counting prediction of $\ra=2.4$. Qualitatively, the transition is very
similar to that observed for randomly diluted 2D CF networks: it is still
second-order, with a cusp in the second derivative of $F$ at the transition, and
the rigidity and stress transitions still coincide.

Self-organization for glass networks is done in the same way as for 2D CF
networks, by rejecting bonds that cause stress. One issue, however, is that for
large network sizes, it
is impossible to create a completely stress-free network by the restricted
dilution process described above. Networks with $\ra=2$ are guaranteed to be
stress-free; however, the restricted dilution process does not allow to go all
the way down to $\ra=2$ (for Bethe lattices the exact lowest achievable $\ra$ is
2.125~\cite{travthorpe} and for the diamond lattice it is found ``experimentally'' to be
around 2.14 or 2.15). Because of this, it is impossible to get rid of all
redundant constraints in the network and, of course, they will stay there when
we start putting the bonds back. The number of these redundant constraints is,
however, very small (about 0.05\% of network constraints are redundant) and,
of course, this number will remain constant as the network is being built if
no new redundant constraints are introduced, so the influence of this problem
is negligible.

\begin{figure}
\begin{center}
\includegraphics[width=8.5cm]{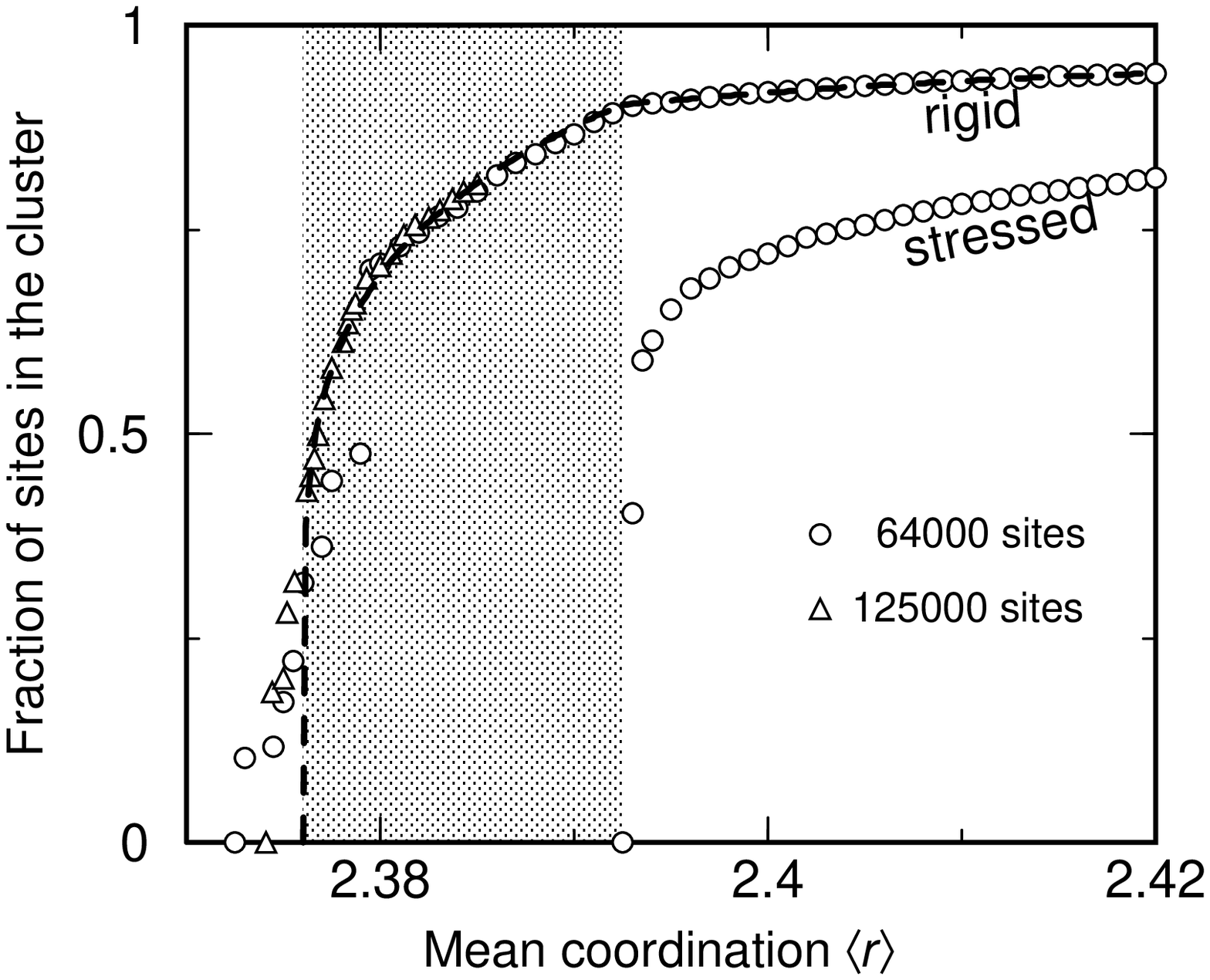}\vspace{0.5cm}
\includegraphics[width=8.5cm]{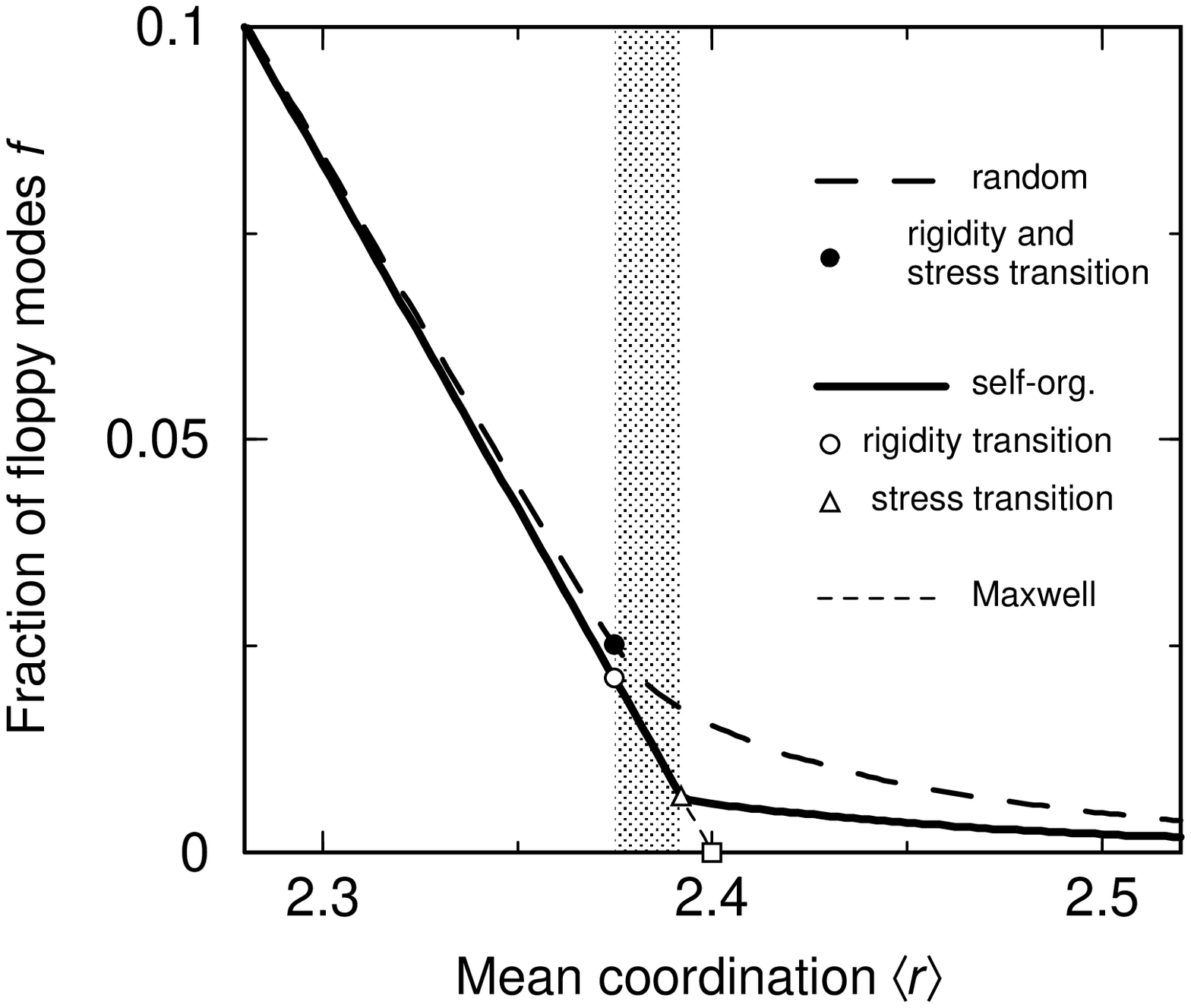}
\caption{The results of rigidity analysis for 3D bond-bending bond-diluted
networks built on the diamond lattice. The figure is adapted from
Refs.~\cite{thorpe00,czech01}. (Top) The fractions of sites in the percolating rigid cluster
and in the percolating stressed region for self-organized networks. Circles
are averages over 4 networks with 64000 sites; triangles are averages over
5 networks with 125000 sites. The dashed line is a power law fit below the
stress transition and for the guidance of the eye above. (Bottom) The number
of floppy modes per degree of freedom, $f=F/3N$, in the random (dashed line)
and self-organized (solid line) cases. Phase transitions are marked, as
specified in the legend. The square is the Maxwell counting value for the
transition, $\ra=2.4$. Unlike in CF networks (Fig.~\ref{2D}), the
number of floppy modes is no longer zero above the stress transition in the
self-organized case. The rigidity transitions in the random and
self-organized cases are no longer at the same $f$ (rather, the values of $\ra$
are close, which is probably coincidental). The intermediate phase in the
self-organized case is shaded in both the top and the bottom panels.\label{3D}}
\end{center}
\end{figure}

The sizes of the percolating rigid cluster and the percolating stressed region
for self-organized 3D BB networks built on the diamond lattice are shown in
the top panel of
Fig.~\ref{3D}. It can be seen that the intermediate phase still exists, spanning
the range from $\ra=2.376$ to 2.392. Note that unlike in the CF case, the
upper limit of the intermediate phase (the stress threshold) is lower than the
Maxwell counting result for the transition ($\ra=2.4$). Also, in the stressed
phase the fraction of the network in the percolating rigid cluster is below 1,
so the network is not completely rigid there. This is also seen in the plot of
the number of floppy modes (bottom panel of Fig.~\ref{3D}), as this number remains nonzero in the
stressed phase. These differences from the CF case are due to 
the fact that there can be several constraints associated with a bond in BB
networks, which gives rise to ``partially redundant'' bonds,
with only a part of the associated constraints redundant while the rest remain
independent changing the number of floppy modes and the configuration of rigid
clusters. Since partially redundant bonds create stress, they have to be
rejected when building self-organized networks. At the point where stressless
insertion becomes impossible, by definition there are no fully independent
bonds, but there can still be partially redundant bonds, whose further insertion
would both create stress and reduce the number of floppy modes. For this reason,
the network does not have to be fully rigid at the point where stress appears.
Just as for 2D CF networks, stress percolates immediately after it appears, so
the stress transition is shifted downward from $\ra=2.4$.
Unlike in CF networks, the numbers of floppy modes at the
rigidity transition in the random and self-organized cases do not have to be the
same, as shown in Fig.~\ref{3D}. Note, though, that the ``partial redundancy''
and its consequences are, in a way, an artifact of the model; there are
variants of the self-organization model with no partial redundancy, one of
which I describe in the next subsection.

Finally, note that from Fig.~\ref{3D}, the rise of the percolating stressed
region size from zero is rather sharp. This may mean that the stress transition
in this case is actually {\it first order}, although the data are also
consistent with a second order transition with a very small critical exponent
($\beta\approx 0.1$). Note that the break in the slope of $F$ is also consistent
with a first order transition, but, as we discussed in the previous subsection,
this break is also present in the case of 2D CF networks, in which case the
transition is almost certainly second order based on the stressed region size.

\subsection{Local structure variability and the width of the intermediate phase}

The self-organization model described here predicts a particular value of the
width of the intermediate phase, $\Delta\ra=0.016$ (from 2.376 to 2.392). On
the other hand, the experimental width varies widely depending on the glass
composition: it can be as high as 0.17 (in
P$_x$Ge$_x$Se$_{1-2x}$~\cite{bool05pgese}),
but can also be essentially zero (in iodine-containing
glasses~\cite{bool01iod}). Ideally, theoretical models should be able to explain this
variation and predict the width for a particular composition. Work by
Sartbaeva {\it et al.}~\cite{asel07}, while not quite achieving that, sheds some light
on possible sources of this variability.

Sartbaeva {\it et al.} considered glass networks with atoms of coordinations
2 (Se) and 4 (Ge) that were built starting from fully 4-coordinated amorphous
networks built using the WWW method~\cite{WWW,djordj} by decorating Ge-Ge bonds with Se
atoms. The number of Se atoms decorating each original Ge-Ge bond and thus
forming a chain is chosen
from a predefined distribution that can be narrower or wider; this distribution
will change when the network is modified as described below. It is argued that
in the body-bar representation of the network (see the section on the pebble
game in this review), each chain of length $l$ (for
$l\le 5$) can be replaced by a bunch of $5-l$ bars (or constraints) between
bodies representing Ge atoms, whereas chains
with $l>5$ can simply be removed without affecting the overall rigidity of
the network. It makes sense then to characterize the distribution of chain
lengths by the variance $v$ of the number of bars associated with each
chain determined as above. The network is then made gradually more rigid by
picking a chain at random and removing an atom from that chain; this increases
the number of effective constraints by one for chains of length 5 or less.
To obtain random networks, all removals are accepted; to model
self-organization, those removals that would create stress are rejected.

Without self-organization, the rigidity and stress transitions coincide, just
like for random bond dilution. Unlike the bond dilution case, though, the
transition can be both first and second order, depending on the constraint
number variability $v$: for low variability, the transition is first-order,
with the jump in the fraction of the network in the percolating cluster almost
equal to one; but at a certain value of $v$, the jump drops to zero and at
higher $v$, the transition is second-order. When self-organization is carried
out, an intermediate phase appears, but only for high $v$, when the transition
without self-organization is second-order; otherwise, the width of the
intermediate phase shrinks to zero. The upper limit of the intermediate phase
(the stress transition) is always at $\ra=2.4$ (since in this model, effectively
one constraint is inserted at a time, as in the 2D CF model); the width of the intermediate
phase is linear in $v-v_c$, where $v_c$ is the threshold at which the
non-self-organized transition switches from first to second order. The
variation in the width of the intermediate phase is thus associated with the
local structural variability, namely, the width of the distribution of chain
lengths. although the maximum width that the authors were able to obtain
(about 0.05) is still way below the maximum value of 0.17 seen experimentally.

Several comments about the work of Sartbaeva {\it et al.} are in order. First,
while it can be proved (essentially rigorously, only assuming the molecular
framework conjecture) that the rigidity transition is indeed first order when
the chain length variability is zero (i.e., all chains are of exactly the same
length)~\cite{chemorder}, this is less obvious in the case of small but nonzero
variabilities. Normally, when a phase transition in a system can be first or
second order depending on the value of some parameter, the threshold value of
the parameter is the so-called tricritical point, with the jump of the order
parameter decreasing gradually to zero as the tricritical point is approached.
No such gradual decrease is observed, with the jump as a function of $v$
changing almost instantaneously from almost 1 to 0 (see Fig.~4 in
Ref.~\cite{asel07}).
As this does not look like
a standard tricritical point, other options remain open, such as a second
order transition with a narrow critical region whose width depends on $v$; the
change of the order parameter over this region is always almost 1, but whether
the critical region appears as infinitely thin (thus looking like a first order
transition) or has a detectable width (thus looking like a continuous transition)
would depend on the resolution [$\sim 1$~bond, or $O(1/N)$ in terms of $\ra$, thus dependent on
the network size]. The second point is that even if we do assume that the
transition is first-order, then unless the jump in the percolating cluster size
is exactly 1 (i.e., the whole network immediately becomes percolating),
the intermediate phase in the self-organized case should still be present,
albeit very narrow: indeed, if at least small pockets of the network remain
floppy at the rigidity transition, a few constraints (a small but finite
fraction) can still be inserted without creating stress. These comments do not
invalidate the essence of the work, of course.

Sartbaeva {\it et al.} also study how the presence of edge-sharing tetrahedra
in the network affects the position of the intermediate phase shifting it
upwards, even beyond 2.4, as observed experimentally in some cases. This effect
was first considered by Micoulaut and Phillips~\cite{micoulaut02,micoulaut03} using a different approach. Of
course, one can ask if edge-sharing tetrahedra should be considered
stressed by themselves and thus excluded from self-organized networks; but in
effect, it is the same type of question as that concerning extra angular
constraints associated with 4-coordinated atoms (which we do allow).

\subsection{Self-organization in connectivity percolation}
An analogous self-organization model can also be considered for connectivity
percolation. When we avoid stress in elastic networks, we avoid
redundancy, and redundancy in connectivity problems means more than one path
connecting two sites, i.e., a {\it loop}. Self-organization then consists in
avoiding loops, by building a network one link at a time but rejecting those
links that close a loop. This is referred to as {\it loopless percolation}. It
is not immediately clear what this means physically, until we recall that
connectivity problems are equivalent to rigidity problems with 1 d.o.f. per
site, and so in systems where for some reason sites are allowed to move in only
one direction avoiding loops means avoiding stress. Even more interestingly,
fully 2D rigidity problems (with 2 d.o.f. per site) are equivalent to
connectivity problems, if both first- and second-neighbor constraints are
present, i.e., for bond-bending networks. Likewise, 3D problems with not just
CF and BB, but also third-neighbor {\it dihedral}, or {\it torsional},
constraints present are also equivalent to connectivity problems. Avoiding
loops in these cases then likewise means avoiding stress, so the results of
connectivity self-organization models may be relevant to systems where
third-neighbor interactions are important.

The history of studies of loopless percolation is rather long and, in fact,
exactly the same model, with bonds inserted one by one and those creating
loops rejected, was proposed first as far back as 1979~\cite{straley79} and rediscovered
again in 1996~\cite{river96}. It was found that percolation occurs before loops become
unavoidable, which in our terms corresponds to the intermediate phase. Namely,
for the square lattice, the lower boundary of the intermediate phase is at
$\ra=1.805$; the upper threshold, at which loops become unavoidable, is,
expectedly, at the Maxwell counting value for the connectivity threshold, which,
for any lattice, is at $\ra=2$. At the point at which any insertion would form
a loop, which corresponds to $\ra=2$ minus one bond, the network is a
{\it spanning tree}. From our perspective, what happens in the ``stressed''
phase is also of obvious interest. Of course, the analog of stressed regions is
now ``loopy'' regions which are contiguous parts of the network consisting of
loops. It turns out that once loops appear, a percolating ``loopy'' region
appears immediately afterwards, so the upper limit of the intermediate phase
can be defined as either the point at which the loops first appear, or,
equivalently, as the point at which they percolate, just as in the rigidity
case.

\begin{figure}
\begin{center}
\includegraphics[width=10.5cm]{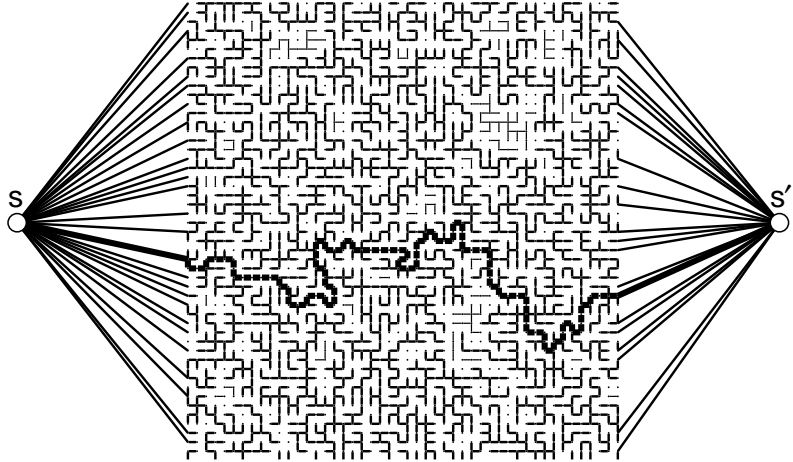}
\caption{A network in the connectivity intermediate phase, on a square lattice
with the source (s) and sink (s$'$) sites added. The thickest bonds represent
the only connection between the source and the sink; the bonds of medium
thickness are other bonds in the percolating cluster (dead ends); the thinnest bonds are
not in the percolating cluster.\label{connect}}
\end{center}
\end{figure}

Just as the elastic moduli in the rigidity case, in the connectivity
self-orga\-nization model the effective conductivity is zero in the thermodynamic
limit in the intermediate phase. This result for connectivity is actually
more obvious than the corresponding rigidity result: because of the absence of
loops, most of the percolating cluster is in dead ends and only sparse filamentary
connections between the opposite sides of the network exist. In
fact, if the boundaries are modified by introducing the ``source'' and ``sink''
sites at opposite boundaries and allowing connections between these sites and
the sites on the respective boundary, there will always
be just one filament between the ``source'' and the ``sink'' (Fig.~\ref{connect}), which obviously makes the conductivity zero in
the thermodynamic limit, especially given that these filaments are fractal
(with the fractal dimension 1.22 in 2D~\cite{middle}) so that their length is much larger than
the linear size of the network. In the ``stressed'' (or ``loopy'') phase, the
result for conductivity is also very interesting: it is exactly linear as a
function of the distance from the stress threshold, $\ra-2$. Note that this
coincides with the effective medium theory result for
conductivity~\cite{kirkpatrick},
whereas in random percolation, the actual result deviates from linearity
close to the transition (in particular, the corresponding critical exponent is
1.30 in 2D~\cite{stauffer}). The exact linearity was proved~\cite{myconduct} in a similar situation,
where bonds were likewise added at random to a spanning tree (just as is done
in this self-organization model); the only difference is that this result is
true with probability 1 for a spanning tree chosen at random, whereas spanning trees
obtained at $\ra=2$ in this self-organization model are biased and actually
belong to the ensemble of so-called {\it minimal spanning trees}~\cite{cormen}.
Nevertheless, numerically, the linearity is at least extremely accurate.
Unfortunately, this linearity is not observed for elastic moduli in the
rigidity case.

\subsection{Stressed but non-stress-percolating intermediate phase}
So far, we have simulated network self-organization by inserting bonds one by
one while trying to avoid stress for as long as possible. What if we invert this
process, i.e., start with the fully coordinated network (obviously, stressed)
and then {\it remove} bonds one by one in a way that {\it gets rid of} stress as
fast as possible? Since removing an unstressed bond does not reduce the amount
of stress in the network, this means removing only stressed bonds. Since within
the topological approach we cannot determine the exact amount by which the
stress energy is reduced when a particular stressed bond is removed, we have
no reason to prefer one stressed bond to another. So in the end, the idea is to
start with the fully coordinated network, pick a bond at random and remove it
if and only if it is stressed (or, alternatively, pick a bond at random from
among those that are stressed and remove it); continue this until no stressed
bonds remain, at which point switch to completely random removal. In cases where
there are several constraints per bond (i.e., BB networks), there is some
ambiguity as to what to do with ``partially stressed'' bonds (i.e., those whose
removal reduces the number of redundant constraints by less than the number of
associated constraints); for simplicity, I will only consider the case when
there is one constraint per bond (CF networks). Note that the procedure is
more computationally expensive than the bond insertion procedure,
as in general there is no easy way to treat bond removal in the pebble game, so
every new network has to be analyzed starting from scratch, even though it
only differs from the previous one by a single removed bond.

First of all, as removal of stressed (i.e., redundant) constraints does not
change the number of floppy modes, this number remains at zero (more properly,
the number of rigid body motions). The number of floppy modes cannot be
smaller than its Maxwell counting approximation, $\fm$; since $\fm$ is above
zero for $\ra<\ra_c$, where $\ra_c$ is the Maxwell counting rigidity threshold
(2 for connectivity, 4 for 2D rigidity), then $\ra_c$ is the point at which
removal of only stressed constraints becomes impossible, since at this point no
stressed constraints are left. At this point, the network is rigid but
stress-free. What happens upon further dilution (which should proceed at random)
is most easily seen using the connectivity example. Since there is no stress
(i.e., loops), the network should look like that in Fig.~\ref{connect}, i.e., it is a tree and
there are only sparse filamentary connections (perhaps as few as one)
between the opposite sides of the
network. Since bonds are only removed and never inserted, it is obvious that
it only takes the removal of an infinitesimally small fraction of network bonds
to destroy all the connections. In other words, the intermediate phase width
shrinks to zero. In the rigidity case, the result is the same.

However, in this model another intermediate phase appears, this time {\it above}
the Maxwell counting transition. To see this, compare the unrestricted bond
removal and the self-organization process using the same random sequence of
bonds for removal but retaining those of the bonds that do not cause stress.
Since removal of unstressed bonds (which is where the two processes differ)
does not affect either the configuration of stressed regions or the number of
redundant constraints, there is a correspondence: random and self-organized
networks having the same number of redundant constraints will have the same
configuration of stressed regions, and in particular, stress either
percolates in both cases or does not percolate in both cases (compare the
similar correspondence for rigid clusters in the case of bond addition). Note, however,
that in the random case some stress exists even below the percolation
transition, so the number of redundant constraints is nonzero even when
stress does not percolate. Because of the above-mentioned equivalence, this
should be true for the self-organized network as well, and so there should be a 
region in the self-organized case where there are redundant constraints (i.e., 
stress), but no stress percolation. This is the new intermediate phase where 
stress occurs but does not percolate. The new intermediate phase
is located {\it above} the Maxwell counting transition, which can be another 
explanation for the experimental observation that the intermediate phase is
sometimes located above $\ra=2.4$. Note also that while the phase with
localized nonpercolating stressed regions came about as a simple consequence
of trying to get rid of stress as quickly as possible, small localized
overconstrained regions can also be energetically favorable compared to
percolating regions for yet another reason: it would not take long for such
small regions to rearrange and turn into nanocrystallites losing stress 
altogether (the purely topological constraint counting would still show the
presence of stress, of course, even when, in a nongeneric nanocrystallite,
there is none). Interestingly, just as in the stress-free intermediate phase,
the elastic moduli (or the conductivity in the connectivity case) should still
be zero in the thermodynamic limit, as the external strain is applied to the
marginally rigid percolating isostatic region, and the more rigid stressed
inclusions do not matter.

The existence of the non-stress-percolating intermediate phase has been
confirmed numerically for the connectivity percolation problem on the square
lattice~\cite{myconduct}. Since the above considerations, strictly speaking, only apply to
the case where there is one constraint per bond, its existence still needs to
be confirmed for bond-bending glass networks, although its absence would be
very surprising.

\sect{Self-organization with equilibration}
Comparing two models considered in the previous section, with self-orga\-nized
networks obtained by bond insertion and bond removal, respectively, we see
that the results differ significantly, even qualitatively, as different kinds
of intermediate phases are obtained. In other words, the results are
history-dependent: they depend on whether the ensemble of networks with a
particular $\ra$ is obtained by assembling the network or by disassembling it.
This is not surprising: the bond insertion algorithm, in particular, resembles
an aggregation process, which is not expected to lead to well-relaxed,
history-independent, equilibrium structures. Indeed, the bond insertion
process below the stress transition disfavors more floppy networks with
smaller rigid clusters: such networks can only be obtained from other networks
with small rigid clusters (as clusters can only grow upon insertion), but such
networks have more places to put a bond without causing stress than more
rigid ones and thus each variant of bond insertion carries relatively less
weight. For similar reasons, the bond removal process above the stress 
transition is biased towards networks with smaller stressed regions. As there
is no reason to prefer any particular stress-free (or minimally stressed)
networks within our approach, the goal should be to generate all such networks
with equal probability, i.e., to obtain the {\it uniform ensemble} of networks.
This section describes the algorithm that can be used to do this and the
results obtained using this algorithm, the most interesting of which is the
existence of a yet different kind of intermediate phase.

\subsection{The equilibration algorithm}
In the connectivity case, the problem of generating the uniform ensemble of
loopless networks has been of interest for a long time due to
the fact that this corresponds to a formal $s\to 0$ limit of the $s$-state
Potts model~\cite{stephen}. Braswell {\it et al.}, in their computational study of
loopless networks~\cite{family}, used the following algorithm to generate the uniform
ensemble. Start with an arbitrary loopless network having a desired number of
bonds (or mean coordination number). Pick a bond at random and remove it, then
reinsert at a place chosen at random from among those where it would not
form a loop. I will refer to this sequence of one bond removal and following
reinsertion as a single {\it equilibration step}. Braswell {\it et al.} showed
that after many such equilibration steps, any loopless network can appear with
the same probability; further equilibration steps would then generate the
uniform ensemble of loopless networks. The proof is based on comparing the
probability of going from an arbitrary network A to an arbitrary network B
reachable in a single equilibration step and comparing with the probability of
going back from B to A; these probabilities, as it turns out, are always equal
and this detailed balance guarantees that in the stationary state, after
reaching equilibrium, probabilities of all networks are the same. The algorithm
can be used in the rigidity case as well (with obvious replacement of
``loopless'' by ``stressless'') and the proof is fully applicable in the
rigidity case as well.

Given the above equilibration procedure, the following algorithm was used
to study self-organization with equilibration~\cite{chubynsky06,briere07}. As in the original
self-organization algorithm, start with an ``empty'' network and add bonds one
by one rejecting bonds creating stress. However, after every bond insertion,
``equilibrate'' the network by doing a fixed number of equilibration steps as
described above. The number of equilibration steps after every bond insertion
should be sufficient for the system to stay equilibrated at all times, in other
words, it should be high enough that increasing this number further would not lead
to a significant change in results. So far, only the stress-free phases have
been studied: once stress becomes inevitable, the procedure is stopped. Ways to
extend the approach to the stressed phase(s) are discussed in the next section.
It is worth noting that even though {\it in general}, as mentioned, it is 
difficult to handle removal of a constraint within the pebble game approach,
the situation is simplified considerably if the constraint being removed is
unstressed: in that case, all one needs to do is release the pebble covering
the constraint. So the algorithm is still rather efficient computationally,
although much less so than the original one without equilibration.

The approach considered here is, in fact, just the zero-temperature limit of
that of Barr\'e {\it et al.}~\cite{barre,barre2}. These authors introduced a model in which
stress was allowed, but there was an associated energy cost proportional to
the number of redundant constraints. The canonical ensemble of networks at a
particular temperature $T$ was considered. In the limit $T\to 0$ (or when the
energy cost per redundant constraint tends to infinity), stress cannot appear
and all stress-free networks are equiprobable, just like in the model
considered in this section. However, Barr\'e {\it et al.} applied their
approach to ``pathological'' Bethe lattices that have an underlying first order
rigidity transition, as opposed to the second order transition in regular
2D CF and 3D BB glass networks.

Chubynsky {\it et al.}~\cite{chubynsky06} and Bri\`ere {\it  et al.}~\cite{briere07} have instead
applied the self-organization with equilibration approach to networks built on
the triangular lattice in 2D. This is expected to be a much better model of
glasses than the Bethe lattice, yet finite-size effects should be less severe
than in 3D, since networks of much larger linear size can be studied with the
same computational effort. The duration of equilibration between bond
insertions was chosen equal to 100 steps above $\ra=3.5$ and 10 steps below,
where stress is rare even when bonds are inserted at random; it was checked that
this is sufficient for convergence.

\subsection{The self-organized critical intermediate phase}
\begin{figure}
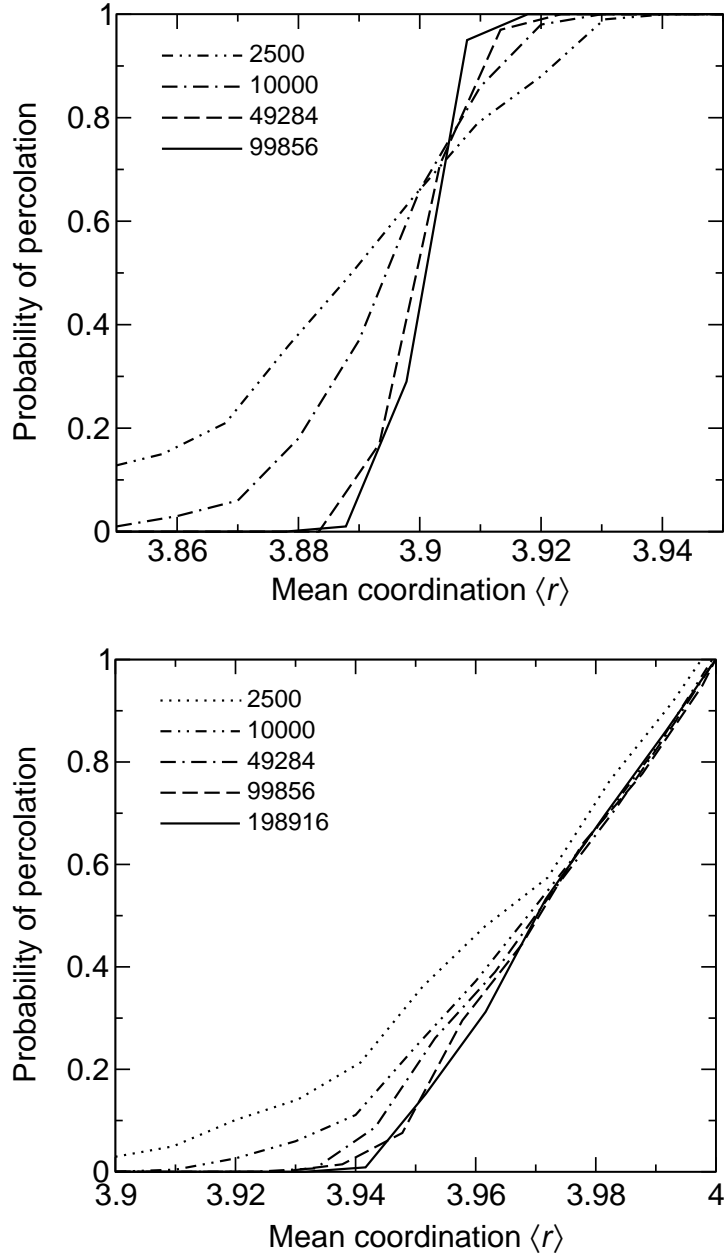

\begin{center}
\includegraphics[width=9.5cm]{figs/percprob_noneq.eps}\vspace{0.5cm}
\includegraphics[width=9.5cm]{figs/percprob_eq.eps}
\caption{The fraction of percolating networks as a function of the mean
coordination number $\ra$, for self-organized networks without
equilibration (top) and with equilibration (bottom) on the triangular lattice.
The network sizes are specified. Without equilibration,
this fraction approaches a step function as the network size increases. With
equilibration, it approaches what appears to be a straight line going from
0 at $\ra\approx 3.945$ to 1 at $\ra=4$, indicating an intermediate phase
between these points. Each curve is obtained from 100 separate runs; in the equilibrated case all networks obtained during
equilibration are used in the analysis. Taken from Ref.~\cite{chubynsky06}.\label{properc}}
\end{center}
\end{figure}
First of all, the study of Chubynsky {\it et al.}~\cite{chubynsky06} confirmed that the
intermediate phase is still
present, however, its properties are very different compared to the model without
equilibration. Normally, percolation thresholds are infinitely sharp in the
thermodynamic limit: if one plots the probability of percolation (or the
fraction of percolating networks) for a given $\ra$, this becomes more and more
similar to the step function as the network size increases, and in the
thermodynamic limit, all networks with $\ra$ below the percolation threshold are
nonpercolating and all networks above the same threshold are percolating. This
is the case not only for random bond insertion, but in the self-organized case
without equilibration as well (top panel of Fig.~\ref{properc}). However, when equilibration is done,
the result is quite different: there is a range of $\ra$ of finite width where,
as the network size increases, the percolation probability approaches a value
between zero and one (bottom panel of Fig.~\ref{properc}). In other words, in
this region both percolating
and nonpercolating networks coexist even in the thermodynamic limit. The lower
boundary of this region lies at $\ra\approx 3.945$. The upper
boundary is at $\ra=4$, i.e., at the stress transition, and so
the intermediate phase of the kind observed without equilibration, where there
is no stress but rigidity {\it definitely} percolates is absent when equilibration
is done. Instead, the region of coexistence of percolating and nonpercolating
networks can be interpreted as a different kind of intermediate phase where stress
is absent and rigidity percolates with a finite probability. Numerically, the
fraction of percolating networks seems to depend {\it linearly} on $\ra$,
growing from 0 at the lower boundary to 1 at the upper boundary, although there
is no proof to date of this linearity. The same kind of
intermediate phase was found by Barr\'e {\it et al.}~\cite{barre,barre2} in 
their Bethe lattice study.

\begin{figure}
\begin{center}
\includegraphics[width=11cm]{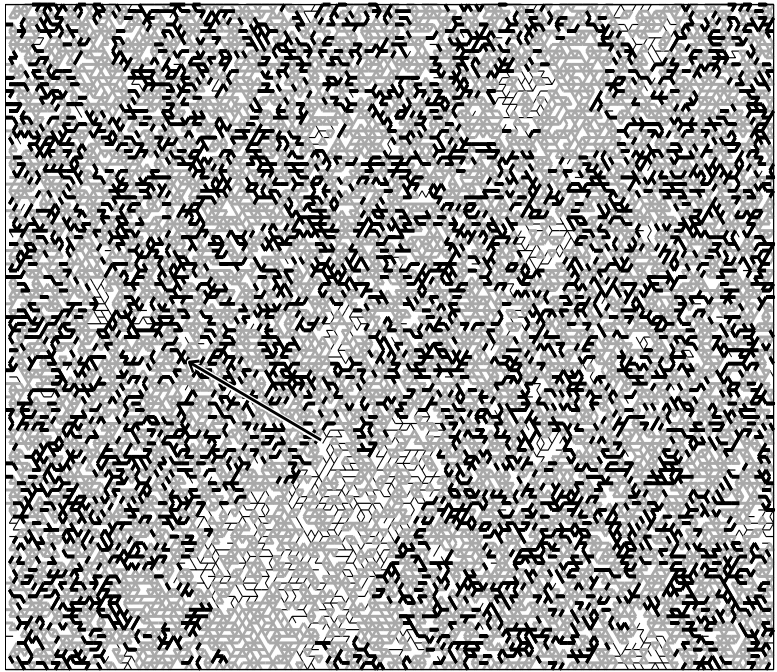}
\caption{An example illustrating how insertion of a single bond (pointed with
an arrow) can influence
the rigidity of a network in the intermediate phase. Thick black bonds are
converted from ``floppy'' to ``rigid''; they are located throughout a
percolating region that takes up much of the network and becomes a single
percolating rigid cluster upon insertion of the bond. Thin bonds remain ``floppy'';
thick gray bonds were ``rigid'' before insertion and remain ``rigid''. The
concepts of ``floppy'' and ``rigid'' bonds are defined in the text. Adapted
from Ref.~\cite{briere07}.\label{conv}}
\end{center}
\end{figure}

\begin{figure}
\begin{center}
\includegraphics[width=9cm]{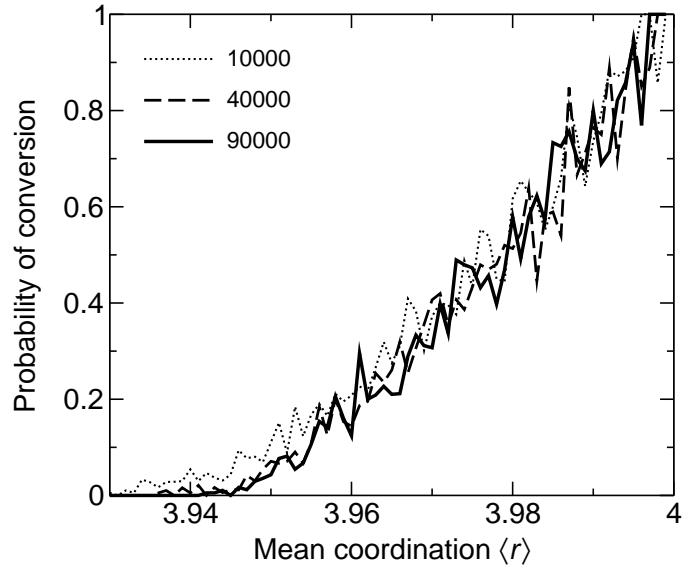}
\caption{The probability that a single bond inserted at a randomly chosen place
where it does not cause stress into a randomly chosen nonpercolating network
converts that network into percolating, for the model of self-organization
with equilibration on the triangular lattice, for different lattice sizes specified
in the legend. Adapted from Ref.~\cite{briere07}.\label{probconv}}
\end{center}
\end{figure}

Not only are both nonpercolating and percolating networks present in the
intermediate phase, it is very easy to switch between the two classes. All it
often takes is insertion or removal of a single bond, which can be viewed as a
microscopic perturbation. The changes in rigidity properties caused by such a
perturbation are often very dramatic. To characterize these changes, it is
convenient to introduce the concept of ``floppy'' and ``rigid'' bonds. It
turns out that in nonpercolating networks in both the floppy phase and the
intermediate phase many rigid clusters contain just a single bond.
Let us call bonds belonging to such clusters {\it floppy} and all other bonds
{\it rigid}. Interestingly, even close to the stress transition, where very
few floppy modes remain, in those few networks that remain nonpercolating
about 1/4 of the bonds are floppy. Insertion of a bond in the network in a place
where it is not redundant and does not cause stress rigidifies the network and
may convert some floppy bonds into rigid. Both the number of such converted
bonds and their location (whether they are all located in the same part of the
network or are spread over the whole network) are good indicators of the
influence of the inserted bond. Fig.~\ref{conv} gives one example of this when a bond
is added to a nonpercolating network. Many bonds are converted from floppy to
rigid and such bonds are present everywhere in a region that spans the network
and takes up most of it. Thus many single-bond rigid clusters (and other small
clusters) merge into a single percolating cluster and the network is converted
from nonpercolating to percolating. Note that this phenomenon is specific to
rigidity, as in connectivity percolation insertion of a single link can at most
join two clusters, but cannot merge more than two clusters. The
probability that a single bond
inserted at a random place (where it is not redundant) into a randomly chosen
nonpercolating network causes percolation is plotted in Fig.~\ref{probconv} for different
network sizes and based on these data, it seems to remain finite even in the
thermodynamic limit inside the intermediate phase. (By the way, the fraction of
nonpercolating networks such that there exist places where bond insertion would
cause conversion nonpercolating$\to$percolating is equal to the probability
of percolation, which can be proved~\cite{briere07}). The outcome of bond removal
from a percolating network in the intermediate phase can be equally
dramatic, with a single bond removal often sufficient to destroy percolation.
In fact, there is a relation between the average numbers of bonds converted
from floppy to rigid upon bond insertion and those converted from rigid to
floppy upon bond removal, as discussed in Ref.~\cite{briere07}.

\begin{figure}
\begin{center}
\includegraphics[width=9cm]{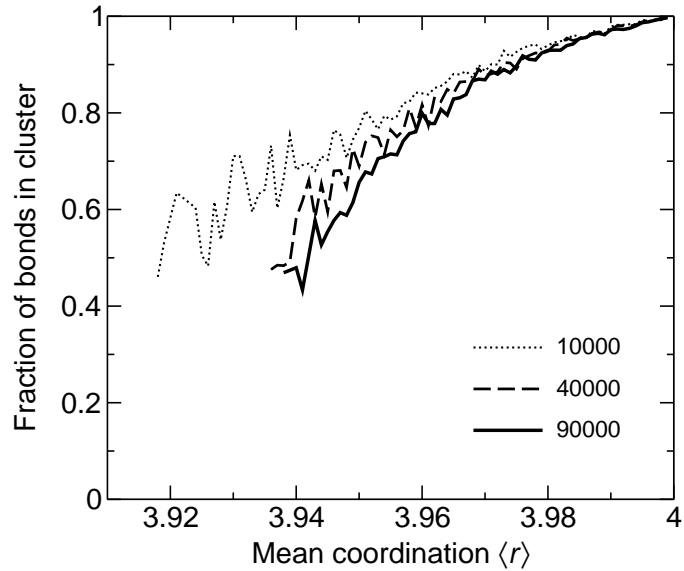}
\caption{The fraction of bonds belonging to the percolating rigid cluster
averaged over all {\rm percolating} networks, for the model of self-organization with equilibration on the
triangular lattice, for different lattice sizes specified in the legend. For each
data point, 100 independent networks were generated and those of them with a
percolating cluster were chosen. At lower $\ra$, the noise in the data is
higher, as there are fewer percolating networks. Adapted from Ref.~\cite{briere07}.\label{percsize}}
\end{center}
\end{figure}

The size of the percolating region formed upon a single bond insertion is also
fairly typical in Fig.~\ref{conv}, where it takes up perhaps about 80\% of the network.
First of all, it was checked that percolating clusters formed as a result of
insertion of a bond into a nonpercolating network have the same size on
average as clusters that were initially percolating. This average percolating
cluster size is plotted in Fig.~\ref{percsize}. It is seen that it is quite large and,
moreover, it appears that it remains nonzero at the lower threshold, as in
first order transitions (although this is not a perfect analogy, as we will
see) . As for the width of the
distribution of percolating cluster sizes (see Fig.~2 in Ref.~\cite{briere07}), it
turns out that this width is small (in other words, all, or at least the vast
majority, of networks have either a large percolating cluster or no percolating
cluster at all); on the other hand, interestingly, this width, while remaining
small, does not go to zero in the thermodynamic limit, in other words, the
percolating cluster size is not a self-averaging quantity.

\begin{figure}
\begin{center}
\includegraphics[width=10.5cm]{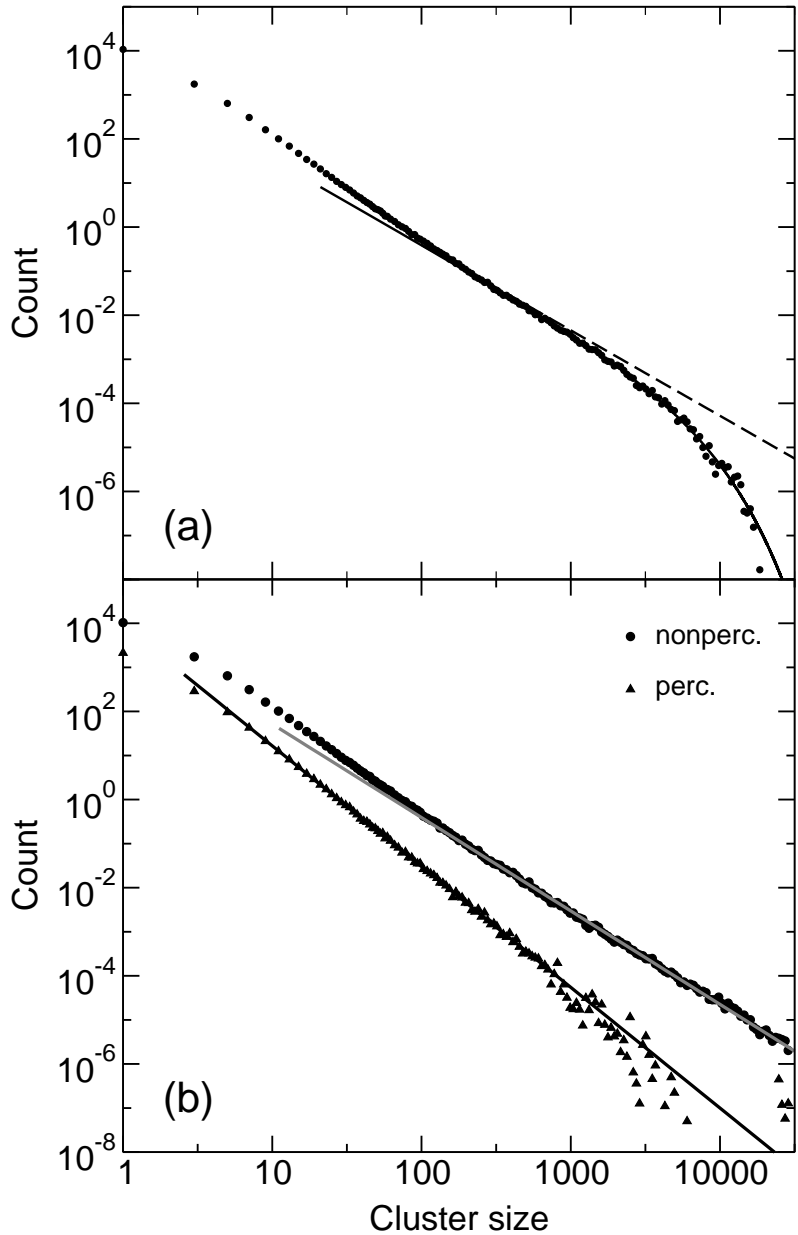}
\caption{The distribution of sizes (given in terms of bonds) of nonpercolating
rigid clusters in nonpercolating networks in the floppy phase at $\ra=3.92$
[panel (a)] and in both percolating and nonpercolating networks in the
intermediate phase at $\ra=3.97$ [panel (b)] in the model of self-organization
with equilibration on the triangular lattice. In panel (a) the solid line is
the fit by a product of a power law and an exponential; the dashed line is the
same power law without the exponential. In panel (b) the lines are power law
fits. The figure is adapted from Ref.~\cite{briere07}; for more details on these plots,
see that reference.\label{clstdist}}
\end{center}
\end{figure}

It is also
interesting to look at the properties of {\it nonpercolating} clusters. For a
second order transition, close to the transition point, the distribution of
nonpercolating cluster sizes is power-law with an exponential cutoff
(corresponding to the correlation radius) that
moves to higher cluster sizes as the transition is approached from either side,
until right at the transition, the power law persists to infinity. For a first
order transition, the crossover to exponential behavior should exist even exactly
at the transition. What is seen in the system considered here is neither of
these two situations. In Fig.~\ref{clstdist}, the cluster size distributions are given for
a point in the floppy phase (upper panel) and a point in the intermediate phase
(lower panel; separately for percolating and nonpercolating networks). The
points are chosen at about the same distance from the transition separating the
phases and thus in a regular second order transition,
the crossover sizes from the power law to the exponential should be about the
same (in the first-order case, both crossover sizes should be rather low, as
there is no divergence of the correlation radius). Instead, while in the floppy
phase the crossover is clearly seen (upper panel of Fig.~\ref{clstdist}), there is no clear
evidence of such a crossover in the intermediate phase (lower panel of
Fig.~\ref{clstdist}), especially for nonpercolating
networks (for percolating ones, statistics is not as good for the largest cluster
sizes as much of the network is taken up by the percolating cluster). That is,
even deep inside the intermediate phase, the cluster size distribution looks as
it normally would exactly at the second order transition point. This, and the
fact that both percolating and nonpercolating networks coexist in the
intermediate phase, indicates that the system self-organizes in such a way as to
maintain itself in a critical state on the rigid-floppy boundary, right on
the verge of rigidity where even a single bond insertion or removal can make
a nonpercolating network percolating or vice versa, not just at a
single value of $\ra$, but everywhere in the intermediate phase. This is
exactly the definition of {\it self-organized criticality}~\cite{bak}, except this
phenomenon is normally observed in out-of-equilibrium systems, whereas here it
is observed in an equilibrated system. Of course, this critical state is not
the same as that in non-self-organized rigidity percolation, since here the
percolating cluster, when it exists, takes up a finite fraction of the network,
whereas in usual rigidity percolation (and connectivity percolation, for that
matter) the percolating cluster emerging at the transition is fractal and thus
occupies an infinitesimally small fraction of the network.

Above, we have considered the effects of insertion of a bond in a nonpercolating
network, in a place where it would not create stress. It is also interesting to
look at what happens when a {\it percolating} network is taken and a bond is
inserted in such a way that a stressed region {\it is} created. It turns out
that this stressed region often percolates and when it does, it takes up a
finite fraction of the network. In other words, in this case the effects of a
microscopic perturbation are again macroscopic.

\subsection{Entropy cost of self-organization}
The self-organization procedure, as it is described here, is designed to
reduce the energy of the network by avoiding unnecessary stress. However, since
glasses form at a nonzero temperature, the {\it entropy} of the system matters
as well. If it turns out that the entropy of the self-organized network is
much smaller than that of the corresponding random network, then in practice,
very little self-organization should take place. This is why estimating the
entropy cost of self-organization is important.

The entropy of a covalent network glass can be viewed as consisting of two
parts, the {\it topological entropy} and the {\it flexibility entropy}. The
topological entropy $S_t$ is related to the number of network topologies
(or bond configurations) corresponding to the thermodynamical state of
interest. The flexibility entropy depends on the configuration space accessed
by each such bond configuration or, crudely speaking, on how flexible the
network is. This division is similar to traditional division into the
configurational and vibrational entropy in the inherent structure
formalism~\cite{stillinger},
although the difference is that a single network may correspond to several
different energy minima (or inherent structures) and thus the flexibility
entropy may include contributions from several inherent structures. Obviously,
the flexibility entropy depends on the details of the potential energy function
and cannot be evaluated within a purely topological approach. Qualitatively,
the flexibility entropy is expected to be roughly proportional to the number of
floppy modes~\cite{jacobsPRE,naumis,thorpe} and since self-organized networks have no redundant constraints
and thus fewer floppy modes than random networks, the flexibility entropy of
the former should be smaller than that of the latter. This difference is
probably not very large, as in real systems the range of floppy motions is
restricted by steric effects, but it certainly needs to be evaluated in the
future using a reasonable potential energy function. On the other hand, assuming
(as we have done so far) that all allowed bond configurations are
equiprobable, the topological entropy for self-organized networks with $N_B$
bonds is simply
\begin{equation}
S_t(N_B)=\ln N_{\rm bc}(N_B),\label{topoent}
\end{equation}
where $N_{\rm bc}(N_B)$ is the number of allowed bond configurations for
networks with $N_B$ bonds and the Boltzmann
constant $k_B=1$. The topological entropy cost of self-organization comes from
restricting the set of allowed networks to those without stress and its
calculation is reduced to calculating the number of stress-free networks having
a given number of bonds (or $\ra$).

The calculation of $N_{\rm bc}$ is based on the following argument. Suppose that
on average, in a stress-free network with $N_B$ bonds there are $n_+(N_B)$ places where
insertion of a bond would not create stress. Then for a network with $N_B$
bonds we will have a list of (on average) $n_+(N_B)$ stress-free networks with $N_B+1$
bonds that can be obtained from this network by insertion of a bond. If all these lists
for each of the $N_{\rm bc}(N_B)$ networks with $N_B$ bonds are merged,
the resulting combined list will contain $n_+(N_B)N_{\rm bc}(N_B)$ entries. But the
number of {\it distinct} networks with $N_B+1$ bonds is smaller: indeed, each
distinct network will be encountered $N_B+1$ times in the combined list. This is
because if we take a network with $N_B+1$ bonds and remove any one of its bonds, this 
will produce a distinct network with $N_B$ bonds from which the original network
with $N_B+1$ bonds can be obtained by bond insertion. Then the number of
distinct stress-free networks with $N_B+1$ bonds is
\begin{equation}
N_{\rm bc}(N_B+1)=N_{\rm bc}(N_B)\frac{n_+(N_B)}{N_B+1}.
\end{equation}

Using Eq.~(\ref{topoent}),
\begin{equation}
S_t(N_B+1)-S_t(N_B)=\ln\frac{n_+(N_B)}{N_B+1}.\label{iter}
\end{equation}
Since there is only one realization with 0 bonds, $S_t(0)=1$.
Using this initial condition, Eq.~(\ref{iter}) can be iterated to obtain
$S_t(N_B)$ for any $N_B$, if $n_+(N_B)$ is known. This latter quantity can be
estimated by a Monte Carlo procedure, by trying to insert a bond in random
places and finding the fraction of attempts that do not create
stress, $\nu(N_B)$; the product of this fraction and the total number of places
where a bond can be inserted, $n_+^{(0)}$, gives $n_+(N_B)$:
\begin{equation}
n_+(N_B)=n_+^{(0)}\nu(N_B).\label{nplus}
\end{equation}
On the other hand, without self-organization all networks are allowed, so for
random networks the topological entropy is
\begin{equation}
S_t^r(N_B)=\ln N_{\rm bc}^r(N_B),
\end{equation}
where the superscript $r$ stands for ``random'' and $N_{\rm bc}^r(N_B)$ is the
number of networks with $N_B$ bonds. This number of networks is, of course, easy
to calculate, as it is just the binomial coefficient giving the number of ways to
choose $N_B$ places to put bonds out of $zN/2$ available places (where $z$ is
the full coordination of the lattice and $N$ is the number of sites). It is,
however, more convenient to obtain an iterative expression for
$S_t^r(N_B)$ similar to Eq.~(\ref{iter}). For this, $n_+(N_B)$ in Eq.~(\ref{iter})
needs to be replaced by the total number of available places to put a bond,
denoted above $n_+^{(0)}(N_B)$:
\begin{equation}
S_t^r(N_B+1)-S_t^r(N_B)=\ln\frac{n_+^{(0)}(N_B)}{N_B+1}.\label{iterran}
\end{equation}
The topological entropy cost of self-organization is
$\Delta S_t(N_B)=S_t^r(N_B)-S_t(N_B)$ and, using Eqs.~(\ref{iter}),
(\ref{iterran}) and (\ref{nplus}), we get
\begin{equation}
\Delta S_t(N_B+1)-\Delta S_t(N_B)=-\ln \nu(N_B).
\end{equation}
Iterating this equation using $\Delta S_t(0)=0$ as the initial condition and
obtaining $\nu(N_B)$ by the Monte Carlo procedure described above,
$\Delta S_t(N_B)$ can be obtained for any $N_B$.

The result of the above procedure is that the topological entropy cost of
self-organization is the highest at the upper boundary of the intermediate phase
($\ra=4$), but even at that point it is only about 2.3\% of the random network
entropy~\cite{chubynsky06}. This means that self-organization is likely to be a significant
factor and the existence of the intermediate phase of the kind described here
is plausible. Of course, this assumes that the flexibility entropy cost that
still needs to be evaluated and depends on the details of the potential is not
too high.

\subsection{Discussion}
A tacit assumption in rigidity percolation studies has always been that
percolating networks should be very different from nonpercolating networks in
their physical properties, especially when the percolating cluster is not
fractal but rather takes up a finite fraction of the network. The first blow to
this view was the conclusion that the elastic moduli are zero in the
intermediate phase of the model of self-organization without equilibration,
despite the existence of a large percolating rigid cluster. But the most
striking result in this respect is the fact that in the intermediate phase of
the equilibrated model, networks with large percolating clusters not only
coexist with nonpercolating networks, but it often takes just one bond to
switch from nonpercolating to percolating and vice versa. Two infinite networks
differing by just one bond cannot possibly be different physically, can they?
One could also ask:
maybe it is stress percolation, rather than rigidity percolation, that is
important? Indeed, it is the stressed backbone that responds to external strain
and should determine the elastic properties of the network; in intermediate
phases, both with and without equilibration, there is no stress, and when a
constraint across the network is added to simulate external strain, the
stressed region that arises is apparently fractal~\cite{czech01}, which seems to confirm this
consideration. But in the equilibrated model, we have seen that adding a single
bond (even locally, not across the network) that causes stress often creates a
macroscopic stressed region that occupies a finite fraction of the network!

What the above means is at least for the model of self-organization
with equilibration considered here, the distinction between
percolating and nonpercolating networks, whether it is rigidity or stress
percolation that is considered, is physically meaningless. However, it would be
a pity to abandon altogether the percolation approach that has served us so well.
Perhaps the solution is to claim that, in fact, {\it all} networks
in the intermediate phase do indeed percolate, but it is not rigidity or
stress percolation, but rather {\it near-isostaticity percolation}. What this
means is even in nonpercolating networks in the intermediate phase there is,
in fact, a percolating region that is nearly isostatic, i.e., d.o.f. and
constraints balance nearly exactly in this region, with the difference being
$O(1)$. The exact value of this difference, of course, should not matter in the
thermodynamic limit. However, technically, the region is ``truly'' isostatic
only if this difference is exactly equal to 3 (in 2D). If the region is slightly
underconstrained, even by one constraint, so this difference is 4, the pebble
game is unable to detect it directly and distinguish it from a completely
floppy region that is not even close to being isostatic. But this region can be
revealed if a single constraint is added to it so it becomes exactly isostatic.
This is what we have seen when a network switched from nonpercolating to
percolating upon a bond insertion --- in fact, it was percolating in both cases,
but in the first case the percolating region was nearly isostatic, rather than
truly isostatic. In some cases, as we know, a single bond insertion does not
turn the network into percolating (no matter where the insertion occurs), but
this simply means that the difference between d.o.f. and constraints is higher
than 4 [but still supposedly $O(1)$].

The concept of near-isostaticity also helps understand the relation between
network rigidity analysis and experiments on real glasses. Explanations in
terms of rigidity of
many phenomena associated with the intermediate phase, such as the absence of
a threshold in the response of the vibrational mode frequencies to
pressure~\cite{bool05press},
assume that there is a percolating region in the network that is isostatic, so
the constraints and d.o.f. balance throughout this region. The presence in the
intermediate phase of nonpercolating networks consisting mostly of small rigid
clusters seemed to contradict this idea. But the above considerations suggest
that the percolating region still exists in all networks, except in some cases 
it is nearly isostatic, rather than exactly isostatic, which should not matter
physically.

\subsection{Connectivity percolation}
Finally, I describe briefly the results of self-organization with equilibration
in the connectivity case. As mentioned, in this case we have loopless
percolation with all loopless networks equiprobable. This is also a particular
limit of the $s$-state ferromagnetic Potts model~\cite{wu} when the interaction energy is
$-sJ$ ($J$ is a finite constant) and $s\to 0$~\cite{stephen}. The limit $T\to 0$ corresponds to
the spanning tree limit at $\ra=2$. The conclusion of studies using both the
percolation approach~\cite{family} and the Potts model approach~\cite{straley90}
is that {\it in 2D} the critical point corresponds to the spanning tree limit, i.e.,
percolation does not occur until $\ra=2$ and thus
there is no intermediate phase. However,
unusually, the correlation radius depends exponentially on
$1/T$~\cite{treesexpon} and so
even for very large systems will quickly exceed the system size producing an
apparent transition at a nonzero $T$ (i.e., in our terms, an intermediate
phase) whose width will decay very slowly (logarithmically) with the network
size. Estimates show that even for a network with the Avogadro number of sites,
the width of the intermediate phase is still perceptible, about 0.02~\cite{looplessunpub}! On the
other hand, {\it in three dimensions}, the transition
occurs at a finite $T$~\cite{loopless3D}, i.e., the intermediate phase does exist. This
intermediate phase is of the ``old'' type, i.e., with percolation occurring with
probability 1, which is not surprising given that the situation where a single
inserted bond can merge an infinite number of small clusters into one
percolating cluster cannot happen in connectivity.

\sect{Conclusions and outlook}
In this paper, I have reviewed various models of elastic network
self-organi\-zation by avoidance of unnecessary stress that possess
an intermediate phase that is believed to correspond to the intermediate
phase observed experimentally in covalent network glasses. The original
self-organization model by Thorpe {\it et al.}~\cite{thorpe00,czech01} has an intermediate phase
that is rigid (there is a percolating rigid cluster) but stress-free. Another
possibility is an intermediate phase where stress is
present but does not percolate~\cite{myconduct}. Finally, in the model with equilibration,
where all stress-free networks are equiprobable, networks with percolating and
nonpercolating rigidity coexist~\cite{chubynsky06}. These intermediate phases seem very different
at first, but in reality they are the same physically, as in all these cases
the network contains a percolating region that is isostatic or nearly isostatic.
In cases where this region lacks a few bonds (even as few as one) to be
truly isostatic, it will not be detected by the standard rigidity analysis, and
hence the network appears nonpercolating, but this region can sometimes be
revealed by insertion of a bond.

There are several possible directions of future studies of models of the type
considered here. First of all, studies of self-organization with equilibration
have so far only been done below the Maxwell counting threshold, i.e., in the
part of the phase diagram where stress is avoidable. Extending these studies
to the stressed part, by producing the ensemble of networks that, although
stressed, have as few redundant constraints as possible, with all such networks
equiprobable, can also give interesting results. Note that the
intermediate phase in the unstressed part of the phase diagram is present in the unequilibrated
model with bond insertion, but not in the model with bond dilution, as in the
latter case the network is always floppy below the Maxwell counting transition.
The equilibrated result is, in a way, intermediate, as the network apparently
fluctuates between nonpercolating and percolating (although we now know that it
is always percolating in reality, in the near-isostaticity sense). On the
other hand, in the stressed part of the phase diagram we also have an intermediate phase present
in one case (this time for bond dilution) and absent in the other case (for
bond insertion), so the result in the model with equilibration may still be
an intermediate phase that technically fluctuates between stress percolation
and stress nonpercolation (with finite stressed regions always present, of
course), although, as in the unstressed part of the phase diagram, we expect
a percolating region to always exist and be nearly isostatic, only this time
sometimes having slightly {\it more} constraints than a truly isostatic region
would have. These are, of course, just speculations and they need to be tested
in actual simulations. The equilibration process would be similar to that used
in the unstressed part of the phase diagram: the equilibration step would consist of a random
bond removal with reinsertion, with the difference that this reinsertion would
be completely random without restrictions if the removed bond was stressed (as the inserted bond
in this case would inevitably be redundant). As mentioned, removal of stressed
bonds is a major complication for the pebble game algorithm, so the procedure
is expected to be much slower than for stress-free networks.

Another important issue is the extension of the equilibrated model to nonzero
temperatures. As mentioned, Barr\'e {\it et al.}~\cite{barre,barre2} studied a similar
model at $T\ne 0$, but this was for the case of Bethe lattices, where
the underlying rigidity transition is first-order, and it was suggested by
these authors that this underlying first order transition is a necessary
condition for the existence of the equilibrated intermediate phase at
nonzero temperatures, and in cases where the transition is second-order, at
$T\ne 0$ the sharp phase transitions at the boundaries of the intermediate phase
are replaced by crossovers. As Sartbaeva {\it et al.}~\cite{asel07} showed, the transition can
be first-order even on regular networks, but it is exactly in these cases that
the intermediate phase nearly vanishes, even without equilibration! The most
straightforward thing to do would be simply repeating simulations by Barr\'e
{\it et al.} on a regular lattice instead of a Bethe lattice. But even
within the approach of Barr\'e {\it et al.} there is freedom in
choosing the energy penalty function. In the work of Barr\'e {\it et al.}, it is
chosen proportional to the number of redundant constraints, but this is an
arbitrary choice and other forms of the
dependence on this number can be considered. Besides, the energy dependence on other factors should be considered,
for instance, it makes sense to assume that the cost of a redundant constraint
would depend on the size of the stressed region that this constraint belongs to.
Even if with all the modifications it turns out that there are no true
phase transitions at nonzero temperatures, the situation is not catastrophic,
since having crossovers that are sharp enough but not infinitely sharp would
still agree with experiment~\cite{barre2}.

The most important issue is, of course, establishing a more quantitative
correspondence between topological models of the type considered here and experiment
(as well as more realistic simulations). Of course, the first question one
should ask is what properties one should look at to establish this correspondence.
X-ray diffraction experiments can, for instance, detect certain features of
medium-range order; however, the medium-range order in self-organized networks
is not very different from that in random networks, as suggested by the very small
topological entropy cost of self-organization. Note that recent X-ray diffraction and EXAFS
work by Shatnawi {\it et al.}~\cite{shatnawi} failed to detect any singularities or breaks
in slope associated with the boundaries of the intermediate phase in the Ge-Se
glass. Likewise, in a simulation using {\it ab initio} or empirical potentials,
even if one could build a well-relaxed network (which is not easy), it would be
very hard to detect particular structural features associated with
self-organization. A straightforward approach is running the pebble game on the
network, but one should keep in mind that even if, e.g., a large percolating stressed
region is detected, it can still be nearly isostatic, a signature of the
intermediate phase. It is not clear at the moment how to deal with this
problem: near-isostaticity means that it should be enough to remove a few bonds
to get rid of the percolating stressed region, but how many should be removed
and how does one pick the ``right'' bonds?

Perhaps the most serious deficiency of the models covered by this review
is their inability
to explain a large width of the intermediate phase observed in many cases,
which can be as high as $\Delta\ra =0.17$. The width obtained in the original
self-organization model without equilibration is 0.016 (from 2.376 to 2.392)
and equilibration reduces the width further (as we have seen in 2D, but this is
likely to be true in 3D as well). Sartbaeva {\it et al.}~\cite{asel07} showed how this
width can be varied by introducing medium-range order in the network, but they
still did not obtain widths larger than 0.05. Note, however, that the
approach of Micoulaut~\cite{micoulaut02,micoulaut03} produces wider intermediate phases
and some clues can perhaps be obtained from that work. Also note that if the
existence of the intermediate phase in the stressed part of the phase diagram
is confirmed, this will increase the total width of the intermediate phase
somewhat.

\sect{Acknowledgements}
First of all, I would like to thank M.F.~Thorpe for introducing me to the topic
of rigidity percolation and many useful discussions. I am also grateful to my
other collaborators, in particular, N.~Mousseau, with whom I had a useful
discussion regarding this review, as well as D.J.~Jacobs, J.C.~Phillips and
M.-A.~Bri\`ere. I also acknowledge discussions with M.~Micoulaut, J.~Barr\'e
and P.~Boolchand. Finally, my apologies to my current supervisor, G.W.~Slater,
for spending time on writing this review, when I should have been studying DNA
electrophoresis!

\vspace{0.9cm} \textsf{\textbf{References}}

\end{document}